%% file: Causal_mechanism_of_extreme_river_discharges_in_the_upper_Danube_basin_network.tex
\documentclass{statsoc}
\usepackage{color}
\usepackage{url}
\usepackage{amsmath,amsfonts,amssymb,amscd,mathtools,bm,bbm}
\usepackage[strict]{changepage}

\usepackage{epsfig}
\newtheorem{theorem}{Theorem}
\newtheorem{postulate}[theorem]{Postulate}
\usepackage{lscape}
\usepackage{enumitem}

\usepackage[english]{babel}
\usepackage{booktabs}
\usepackage{cases}
\usepackage{eucal}
\usepackage{float}
\usepackage{dsfont}
\usepackage{epsfig}
\usepackage{fontenc}

\usepackage[a4paper]{geometry}

\usepackage{hyperref}

\usepackage{lscape}
\usepackage{mathrsfs}
\usepackage{natbib}
\usepackage{rotating}
\usepackage{subfig}
\usepackage{verbatim}
\usepackage[table]{xcolor}

\input{short_bib.tex}

\newcommand{\alphab}{\ensuremath{\bm\alpha}}
\newcommand{\betab}{\ensuremath{\bm\beta}}

\newcommand\independent{\protect\mathpalette{\protect\independenT}{\perp}}
\def\independenT#1#2{\mathrel{\rlap{$#1#2$}\mkern2mu{#1#2}}}

\defcitealias{BLU2006}{BLU, 2007}  
\defcitealias{NAP21852}{NASEM, 2016}
\defcitealias{hannart2016}{Hannart et al., 2016}
\defcitealias{Naveau2018}{Naveau et al., 2018}

\date{}
\title[Causal mechanism of extreme river discharges]{Causal mechanism of extreme river discharges in the upper Danube basin network}
\author[Linda Mhalla {\it et al.}]{Linda Mhalla}
\coaddress{Linda Mhalla, HEC Montr\'eal, 
3000, chemin de la C\^{o}te-Sainte-Catherine, Montr\'eal (Qu\'ebec),
Canada H3T 2A7.}
\address{Department of Decision Sciences, HEC Montreal, Canada} 
\email{linda.mhalla@hec.ca}
\author{Val\'{e}rie Chavez-Demoulin}
\address{Faculty of Business and Economics, University of Lausanne, Switzerland}
\author[{L.~Mhalla,  V.~Chavez-Demoulin and D. J. Dupuis}]{Debbie J. Dupuis}
\address{Department of Decision Sciences, HEC Montreal, Canada}
\begin{document} 
\begin{abstract}
Extreme hydrological events in the Danube river basin may severely impact human populations, aquatic organisms, and economic activity. One often characterizes the joint structure of the extreme events using the theory of multivariate and spatial extremes and its asymptotically justified models. There is interest however in cascading extreme events and whether one event causes another. In this paper, we argue that an improved understanding of the mechanism underlying severe events is achieved by combining extreme value modelling and causal discovery. We construct a causal inference method relying on the notion of the Kolmogorov complexity of extreme conditional quantiles. Tail quantities are derived using multivariate extreme value models and causal-induced asymmetries in the data are explored through the minimum description length principle. Our CausEV, for Causality for Extreme Values, approach uncovers causal relations between summer extreme river discharges in the upper Danube basin and finds significant causal links between the Danube and its Alpine tributary Lech.
\end{abstract}
.

\section{Introduction} 
\label{intro}
The upper Danube basin is regularly affected by flooding and has
received much attention in the hydrological literature. A wealth of studies focus on understanding the flooding processes from hydrological and anthropogenic perspectives; e.g. \citet{Merz2008,Skublics2016}, while others focus on analysing the influence of flood impact variables on monetary flood damage \citep{Thieken2005} and assessing the flood risk management system in Germany (see \cite{Thieken2016} and references therein).
%where flood-affected households on the Bavarian Danube catchment were investigated.} 
%Extreme water discharges and precipitations lead to major damages,

%\bigskip
%Figure \ref{fig:topo_map} shows the topographic map of the upper
%basin with the sites of 31 gauging stations.

\begin{figure}[!h]
	\centering 
 \includegraphics[width=1\textwidth]{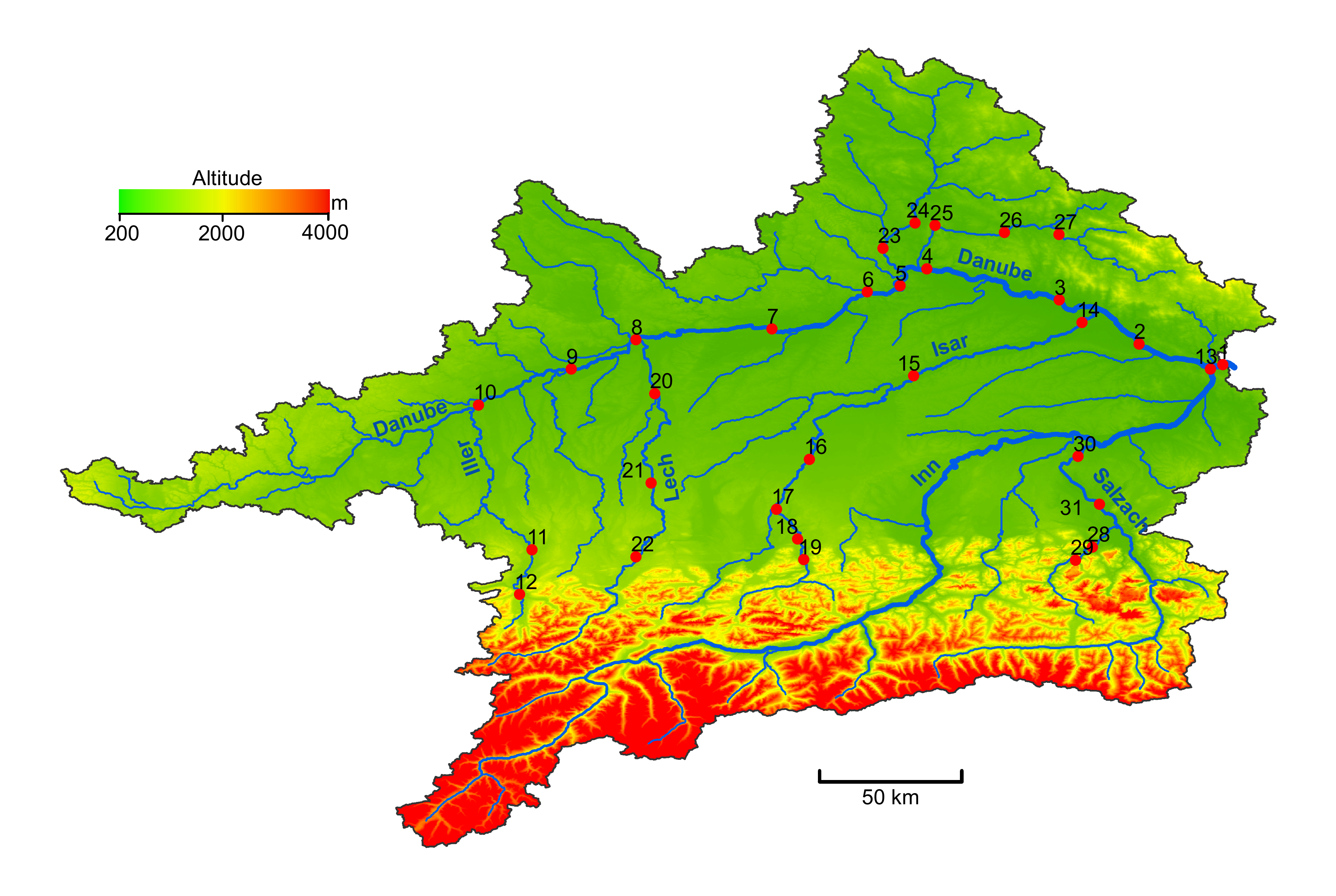}
	\caption{Topographic map of the upper Danube basin, showing sites of the 31 gauging stations along the Danube and its tributaries (courtesy of the authors of \cite{AsadiDavisonEngleke2016}).}
	\label{fig:topo_map}
\end{figure}

\bigskip
Extreme discharges in the upper Danube basin have been studied.
\cite{AsadiDavisonEngleke2016} develop models for spatial extremal dependence based on the hydrological and geographical
properties of a network and apply their methods to the river discharges
at the stations in Figure \ref{fig:topo_map}. Their method allows direct estimation and comparison of the influence of the Euclidean and river distances on the dependence between extreme river discharges. Using a parametric model for multivariate threshold exceedances,
\cite{Engelke_Hitz} develop graphical models for extremes, resulting
in an estimate of an undirected graph structure on the river network
in Figure \ref{fig:topo_map}. Their findings support evidence for the presence of extremal dependence between some flow-unconnected gauging stations due to the spatial extent of extreme precipitation events.

\bigskip
According to the Future Danube Model, an increase of magnitude and
frequency of floods in the Danube basin is expected in 2020--2049,
see \cite{Hattermann2018}. The projected increase in the 
frequency of the $100$-year flood is moreover 
more pronounced in the German part
of the catchment than in the Austrian and Hungarian parts; see, e.g., Figure~6 of \cite{Hattermann2018}. These increases will amplify current
problems in the area and long-term planning would benefit from a
greater understanding of any causal mechanisms in the extremes of river discharges.

\bigskip
Several floods occurred in June 2013, and \cite{Bloschl2013}
analyse the causal factors in terms of atmospheric situation, runoff generation and propagation of the flood wave along the Danube and its tributaries. Although the severity of the floods depends on characteristics such as rainfall duration and soil moisture, an analysis of the spatial causality structure of the severe floods is important for an improved understanding of the environment, a flood risk assessment and management, and an identification of flood mitigation measures. 
%A closer monitoring of the sites that causally affect the network will then be possible.
In this paper, we develop a new method to
assess the intrinsic physical causal relations between 
the extreme discharges in the upper Danube basin network. The effect of time's arrow \citep{Sugimoto2016,Muller2017,Koutsoyiannis2019} on streamflow dynamics will not be considered and the resulting causal mechanisms should be interpreted in conjunction with the physical dynamics of the river network.

\bigskip
Several methods of 
causal inference from observational data at the average level of their
values have been proposed \citep{Maathius2016}. 
While statistical association, possibly occurring without causation,
describes settings where events take place jointly more often than they 
are expected to happen separately, causal relationships allow the 
prediction of the effect of interventions on the observed system.
For instance, climate researchers study causal links between climate 
forcings and observed responses with the aim of attributing likely
causes for a detected climate change \citepalias{NAP21852,hannart2016,
Naveau2018}. If one wants to investigate the effect of
manipulating a variable, such manipulations should be carried out 
in a controlled experiment and the results observed. However,
it is often unethical (clinical studies) or physically impossible
(environmental studies) to perform such experiments, so approaches to discovering causal knowledge based on purely observational
data have been proposed. In contrast with Rubin's causal model and 
the associated counterfactual framework \citep{Rubin1974}, 
these methods are derived from the notion of independence between the cause and the effect conditional on the cause, the causal Markov condition \citep{Spirtes2000}. This notion
of conditional independence, often considered a postulate for causal
conclusions, emanates from the common cause principle of
\cite{Reichenbach1956} known as ``no correlation without causation'' 
and stating that statistical dependence between two random variables
must be due to a causal link where either one causes the other or there 
is a third variable causing both. Moreover, the causal Markov condition 
is at the heart of the theory of causation of \cite{Pearl1995,Pearl2000}
based on structural causal models associated with directed acyclic
graphs. Finally, the Markov condition can be stated in a statistical 
or an algorithmic version \citep{Janzing2010,Lemeire2013} fundamental difference residing in the definition of the mutual
information measuring the violation of conditional independence. 
%\textcolor{green}{Under the same notion of causality, related approaches specifically based on the asymmetry of streamflow and rainfall dynamics have been recently developed \citep{Sugimoto2016, Bardossy2017, Muller2017, Koutsoyiannis2019}.}
In a statistical approach, the mutual information is measured using Shannon entropy while Kolmogorov complexity is used in an algorithmic approach.

\bigskip
In this paper, we are interested in unveiling causal links between
large observed values of two random variables. That is, we want to
know whether an intervention on the tail distribution
of one random variable affects the tail distribution of another random
variable. Unlike bivariate causal discovery methods concerned with causal effects at the mean
level \citep{Peters2014,Marx2018}, our focus is on the conditional
independence in the joint upper tails.
Classical statistical techniques that usually provide a good description
of data tend to fail in the joint upper tail
due to the scarcity of observations.
We turn to extreme value theory for suitable models to describe the observed and unobserved large events
at a penultimate level. The proposed method is valid under the assumption of asymptotic dependence where tail properties are appropriately captured by non-degenerate multivariate extreme value models \citep{deHaan_Book}. 
Inference for the causal structure in the tails relies on the algorithmic version of the causal
Markov condition. The Kolmogorov complexity being non-computable \citep{Cover1991}, we use the minimum description length principle
to provide a well-founded approximation of this complexity, like \cite{Tagasovska2018}.

\bigskip
The rest of the paper is organized as follows. In Section~\ref{evt}
we review basic extreme value theory for the univariate
and multivariate settings, with a focus on threshold methods for joint occurrences of extreme events.
In Section~\ref{causal}
we detail our quantile-based method for distinguishing between
cause and effect at extreme levels of a bivariate random vector. Our CausEV approach is assessed and compared to alternative non-extreme-based methods 
by simulation in Section~\ref{simul}.
In Section~\ref{danube}, CausEV is used to uncover the
causal mechanism between the extreme discharges of the 31 stations 
in Figure \ref{fig:topo_map}. All the directed edges of the 
resulting network coincide with the real flow direction between sites, and causal relationships between summer extreme discharges at Alpine stations and the Danube are uncovered.
We conclude in Section~\ref{conclu}.

\section{\textsf{Extreme value theory}}
\label{evt}
In this section, we present only the main results in extreme value theory (EVT) needed for our development of causal discoveries at extreme levels.
We refer the interested reader to \cite{embrechts1997,Beirlant_2004} 
and \cite{deHaan_Book} for a more comprehensive review of the subject.

\subsection{\textsf{Univariate extreme value theory}}
\subsubsection{\textsf{Maxima of iid random variables}}
\label{sec:block}
Let $(Y_{i})_{i \geq 1}$ be a sequence of independent and
identically distributed (iid) random variables with common distribution $F$.
Let $M_{n}$ be the maximum of a sequence of $n$ such random variables,
i.e.\ $M_{n} = \text{max} \{Y_{1}, \ldots, Y_{n} \}$. 
The Fisher--Tippett theorem \citep{Fisher,Gnedenko} states
that if there exist sequences of constants $\lbrace a_{n}>0 \rbrace$ 
and $\lbrace b_{n} \rbrace$ such that the normalized random variable 
$M_{n}$ converges in distribution to a random variable with 
a non--degenerate distribution function $G$, i.e.,
\begin{equation} \Pr \lbrace (M_{n}-b_{n})/ a_{n} \leq y \rbrace \rightarrow G(y), \quad n\rightarrow \infty ,\label{EVT_thm}\end{equation} then $G$ belongs to the Generalized Extreme Value (GEV) family of distributions
 \begin{align*}
 \text{GEV}_{(\mu,\sigma,\xi)}(y) = \left\{
 \begin{array}{ll}
 \exp\left[ - \left\lbrace 1+ \xi (y- \mu)/\sigma \right\rbrace ^{-1/\xi}_{+} \right], & \xi \neq0, \\
 \exp\left[ -\exp\{-(y-\mu)/\sigma\}\right], & \xi =0,
 \end{array}
\right. \notag
 \end{align*}
defined on $\lbrace y: 1+ \xi (y- \mu)/ \sigma > 0 \rbrace$,
with $-\infty < \mu,\xi < \infty$, $\sigma > 0$, and $x_{+}= \max (x,0)$. 
The parameters $\mu$, $\sigma$, and $\xi$ correspond, respectively, to the location,
scale and shape.
The value of the shape determines the limiting distribution:
Fr\'echet ($\xi >0$) with support bounded below $(\mu - \sigma/\xi, +\infty)$,
reversed Weibull ($\xi < 0$) with support bounded above $(-\infty,\mu - \sigma/\xi)$,
and Gumbel ($\xi=0$) with support in $\mathbb{R}$ and exponential decay in the upper tail.
The Fisher--Tippett theorem provides a justification for the
approximation of the distribution of the maximum in a block of $n$ iid random
variables by the GEV distribution, for sufficiently large $n$. Standard inference techniques include maximum likelihood estimation (MLE) methods where classical asymptotic results of the estimators hold under certain regularity conditions outlined in \cite{Smith} and detailed in \cite{Bucher2017} using the notion of differentiability in quadratic mean.
% Here, the normalizing constants stated in the above result are absorbed by the location and scale parameters of the GEV distribution which is fitted by maximum likelihood methods \cite{coles2001}.

\subsubsection{\textsf{Threshold exceedances}}
\label{sec:threshold}
Extreme events being scarce by definition, the block maximum approach 
may be wasteful, as it discards events that are not as extreme 
as the block maximum but that should be informative about the behaviour
in the tails. An alternative approach to the block maximum is the peaks over threshold where focus is on the asymptotic distribution of the exceedances of a high fixed threshold. The following result \citep{balkema1974}
allows the approximation of the conditional distribution of the
exceedances above a high threshold. If there exist normalizing
sequences $\{a_n>0\}$ and $\{b_n\}$ such that \eqref{EVT_thm} holds,
i.e., $F$ is in the max-domain of attraction of a 
$\text{GEV}_{(\mu,\sigma,\xi)}$, then, for a sufficiently high
threshold $u$ we can model the limiting distribution of the
exceedances $Y-u\vert Y>u$ with a Generalized Pareto Distribution (GPD) 
$G_{(\tilde{\sigma}_u,\xi)}$,
\begin{equation}
\Pr\left\lbrace \left( Y_n-b_n\right) /a_n > u+y \vert \left( Y_n-b_n\right) /a_n >u \right\rbrace \xrightarrow[n \to \infty]{} \left\{
 \begin{array}{ll}
 \left( 1+\xi y/\tilde{\sigma}_u\right) _{+}^{-1/\xi}, & \xi \neq0, \\
 \exp(-y/\tilde{\sigma}_u), & \xi =0,
 \end{array}
\right. \label{gpd_asymptotic}
\end{equation}
where $\tilde{\sigma}_u = \sigma+ \xi(u-\mu)$ and the
shape parameter $\xi$ equals that of the corresponding 
GEV distribution. The limiting distribution function $G_{(\tilde{\sigma}_u,\xi)}$ in \eqref{gpd_asymptotic} is defined on $\{y: y>0 \ \text{and} \ (1+\xi y/\tilde{\sigma}_u)>0\}$,
and the case of $\xi=0$ is interpreted as the limit.
By a slight abuse of notation, we say that 
$Y\mid Y>u\sim \text{GPD}(u,\tilde{\sigma}_u,\xi)$ whenever
$Y-u\mid Y>u \sim G_{(\tilde{\sigma}_u,\xi)}$. That is, $\text{GPD}(u,\tilde{\sigma}_u,\xi)$ is a Generalized Pareto Distribution $G_{(\tilde{\sigma}_u,\xi)}$ shifted by the threshold $u$.

\bigskip
We use the asymptotic result \eqref{gpd_asymptotic} to justify the use of the GPD as the
model for the exceedances $Y_i-u$ for $i$ such that $Y_i>u$ for some
high threshold $u$. The threshold $u$ is chosen following a bias-variance
trade-off: a low threshold will yield a higher number of exceedances and
decrease the variance however it increases 
the bias as the asymptotic approximation 
for the tail of the distribution will be poor.
The GPD may be fitted using MLE \citep{coles2001} or other estimation
methods, see \citet[Section 5.3]{Beirlant_2004}.

\subsection{\textsf{Multivariate extreme value theory}\label{sec:MEVT}} 
\subsubsection{\textsf{Normalized componentwise maxima of iid random vectors}\label{sec:componentwise}}

Let $(\mathbf{Y}_i)_{i \geq 1}$ be iid copies of a $d$-dimensional random vector $\mathbf{Y}=(Y_{1},\ldots,Y_{d})$ with marginal distributions $F_j$, $j=1,\ldots,d$, and joint distribution $F$. We denote by $\mathbf{M}_n=(M_{n,1},\ldots,M_{n,d})$ the vector of componentwise maxima, where $M_{n,j}=\max_{i=1}^{n}Y_{i,j}$ is the sample maximum of the $j$-th component. Note that $\mathbf{M}_n$ is not necessarily observed as the componentwise maxima may occur at different times. As in Section \ref{sec:block}, the vector $\mathbf{M}_n$ needs to be suitably normalized to avoid degeneracy of its limit law as $n \rightarrow \infty$. We suppose that there exist sequences $\{\mathbf{a}_n\} \subset \mathbb{R}^{d}_{+}$ and $\{\mathbf{b}_n\} \subset \mathbb{R}^{d}$ such that the normalized vector $(\mathbf{M}_n-\mathbf{b}_n)/\mathbf{a}_n$ converges in distribution to a random vector $\mathbf{Z}=(Z_1,\ldots,Z_d)$ with joint distribution $G$ and non-degenerate margins $G_j$, $j=1,\ldots,d$. When the marginal distributions $F_j$ are unit Fr\'echet, i.e, $F_j(y)=\exp(-1/y)$, $y>0$, Pickands representation theorem \citep[][Theorem~8.1]{coles2001} states that the law of the standardized componentwise maxima $n^{-1} \mathbf{M}_n$ converges in distribution to a multivariate extreme value distribution (MEVD), $G(\mathbf{z}) = \exp \left\lbrace -V(\mathbf{z}) \right\rbrace,$ with 
\begin{equation}
  V(\mathbf{z}) = \int_{S_d} \max 
  \left( \frac{w_1}{z_1}, \ldots, \frac{w_d}{z_d}\right)  \,d H(\mathbf{w}), \quad \mathbf{z} \in (0, \infty)^d,
 \label{moment_constraint_H}
\end{equation}
for some positive finite measure $H$ on the unit simplex $S_d = \big\lbrace (w_1,\ldots,w_d) \in [0,1]^{d}:\allowbreak w_1+\cdots+w_d =1 \big\rbrace$ obeying
\begin{equation}
  \int_{S_d} w_j \, d H(\mathbf{w}) = 1, \quad j=1,\dots,d. \notag
\end{equation}
The class of MEVDs coincides with the class of max-stable distribution functions with non-degenerate margins \citep[Chapter 8.2.1]{Beirlant_2004}, with a $d$-variate distribution $G$ said to be max-stable if there exist vectors $\alphab_k>\mathbf{0}$ and $\betab_k$ such that
\begin{equation}
G(\alphab_k \mathbf{z} +\betab_k)^k = G(\mathbf{z}), \quad \mathbf{z}=(z_1,\ldots,z_d) \in \mathbb{R}^d, \notag
\end{equation}
for any integer $k >0$. Two results follow from this property. The margins $G_j$ must be max-stable, or equivalently $\text{GEV}_{(\mu_j,\sigma_j,\xi_j)}$, which follows from the univariate EVT and the construction of $\mathbf{M}_n$, and the distribution function $G$ is max-infinitely divisible, i.e., $G^{1/k}$ is a distribution function for any integer $k >0$ \citep{balkema1974}. The max-stability of the limiting distribution $G$ implies that its associated copula,
\begin{eqnarray}
C^{EV}(\mathbf{v}) &=& \exp\left[ -V\left\lbrace -1/\log(\mathbf{v})\right\rbrace \right] , \quad \mathbf{v}=(v_1,\ldots,v_d) \in [0,1]^d, \notag \\
&=& \exp\left\lbrace \left( \sum_{j=1}^d \log v_j\right) A\left( \dfrac{\log v_1}{\sum_{j=1}^d \log v_j},\cdots, \dfrac{\log v_d}{\sum_{j=1}^d \log v_j}\right) \right\rbrace \label{ev_pickands}
\end{eqnarray}
belongs to the large class of extreme value copulas $\mathcal{C}^{EV}$ satisfying
\begin{equation}
C^{EV}(\mathbf{v}) = \left\lbrace C^{EV}\left( v_1^{1/k},\ldots ,v_d^{1/k}\right) \right\rbrace^k, \quad \mathbf{v}=(v_1,\ldots,v_d) \in [0,1]^d, \quad k>0.\label{ev_copula}
\end{equation}
The function $A$ in \eqref{ev_pickands} is termed the Pickands dependence function and is a continuous convex function defined on the unit simplex $S_d$ and satisfying $\max(w_1,\ldots,w_d) \leq A(w_1,\ldots,w_d) \leq 1$, for all $(w_1,\ldots,w_d) \in S_d$. The Pickands dependence function describes the extremal dependence, i.e., the dependence structure in the limiting distribution of the normalized maxima. In the bivariate setting, a useful summary of the strength of tail dependence in $(Y_1,Y_2)$ is given by the coefficient of tail dependence $\chi$ \citep{Coles1999} where
\begin{equation}
\chi_u = \Pr\left\lbrace Y_2> F_2^{-1}(u) \vert  Y_1>F_1^{-1}(u)\right\rbrace \rightarrow \chi \geq 0, \quad u \rightarrow 1. \label{chi_formula}
\end{equation}
When $\chi>0$, the random vector $(Y_1,Y_2)$ is said to be asymptotically dependent, whereas $\chi=0$ characterizes the asymptotic independence regime where the limiting extreme value copula coincides with the independence copula.

\subsubsection{\textsf{Joint threshold exceedances}\label{sec:joint_exc}}
From Section \ref{sec:threshold}, the GPD is suitable for modelling exceedances of a univariate random variable above a high threshold. When the interest is in joint exceedances above high thresholds of a random vector, it is thus reasonable to model their joint distribution with a distribution with GPD margins. When it comes to the dependence structure of this joint limiting distribution, \cite[Section 8.3.2]{Beirlant_2004} and \cite[Section 7.6.1]{mcneil2015} argue that, assuming that the joint distribution of the random vector $\mathbf{Y}=(Y_{1},\ldots,Y_{d})$ is in the maximum domain of attraction of an MEVD, we can approximate, for $\mathbf{Y} \geq \mathbf{u}$ (with inequality holding componentwise), the dependence structure between the exceedances of the high multivariate threshold $\mathbf{u}$ by an extreme value copula satisfying \eqref{ev_copula}.
% Not pertinent: When interest is in threshold exceedances occuring in at least one margin, \cite{Rootzen2018} derived the multivariate Generalized Pareto distribution which is the limiting distribution of the threshold exceedances of a random vector conditional on at least one of its margins exceeding a high threshold.

\subsubsection{\textsf{Inference for extremal dependence}\label{sec:inference_ev_copula}}
As opposed to the univariate case, in which a parametric family of distributions
characterizes all the possible limiting distributions of suitably normalized
maxima, the class of multivariate extreme value distributions yields an infinite-dimensional family of representations. The validity of a
multivariate extreme value distribution relies solely on its associated 
measure $H$ satisfying the mean condition \eqref{moment_constraint_H}. 
For extremal dependence modelling and inference, one can either rely on 
flexible classes of parametric models \cite[Section 9.2.2]{Beirlant_2004}
or use non-parametric estimation of the extreme value copula or its
associated Pickands dependence function.

\bigskip
In this paper, we make no assumption on the form of the extremal 
dependence in the random vector $\mathbf{Y}=(Y_{1},\ldots,Y_{d})$,
and take a non-parametric inference approach. To do so, we use the
min-projection approach of \cite{Mhalla_OT_CD} for inference on the extreme
value copula describing the dependence structure between multivariate 
threshold exceedances, through non-parametric estimation of its
associated Pickands dependence function. We compute the min-projection for a sequence of fixed directions in the unit simplex and regularize the Pickands function so that the resulting estimates of the extreme value copula as well as its derivatives are valid with respect to the convexity and boundary conditions. When the regularisation relies on the median smoothing approach of \cite{Maechler}, the resulting valid estimate of the Pickands function is a linear combination of B-spline basis functions. Equation~\eqref{ev_pickands} implies that the derivatives of the extreme value copula and the Pickands function are related, in the bivariate setting, through
\begin{small}
\begin{eqnarray}
\partial_{v_1} C^{EV} (v_1,v_2) &=& C^{EV}(v_1,v_2) \left\lbrace \dfrac{1}{v_1} A\left( \dfrac{\log v_1}{\log v_1 + \log v_2}\right) + \dfrac{1}{v_1} \dfrac{\log v_2}{\log v_1 + \log v_2} A' \left( \dfrac{\log v_1}{\log v_1 + \log v_2}\right)\right\rbrace,  \notag \\
 \partial_{v_2} C^{EV} (v_1,v_2) &=& C^{EV}(v_1,v_2) \left\lbrace \dfrac{1}{v_2} A\left( \dfrac{\log v_1}{\log v_1 + \log v_2}\right) - \dfrac{1}{v_2} \dfrac{\log v_1}{\log v_1 + \log v_2} A' \left( \dfrac{\log v_1}{\log v_1 + \log v_2}\right)\right\rbrace, \notag
\end{eqnarray}
 \end{small}
where $A(\omega) \equiv A(\omega,1-\omega)$ and $A'(\omega)=d A(\omega)/d \omega$. Thus, inference for the extreme value copula $C^{EV}$ and its partial derivatives is conducted straightforwardly based on the B-spline representation of the Pickands estimator.
\section{\textsf{Pairwise causal discovery of extremes}}
\label{causal}
In this section, we develop a quantile-based method for distinguishing between cause and effect at extreme levels of a bivariate random vector $(X,Y)$. Throughout this section, we assume that we observe a dataset $\{(X_i,Y_i)\}_{i=1}^n$ and that the resulting extreme events are defined as the observations exceeding sufficiently high thresholds in both margins, i.e., the $n_u$ observations $\{(X^{\text{ext}}_i,Y^{\text{ext}}_i)\}_{i=1}^{n_u}$ where for all $i=1,\ldots,n_u$, $X_i^{\text{ext}}>u_X$ and $Y_i^{\text{ext}}>u_Y$ for high thresholds $u_X$ and $u_Y$. We study the causal relationships between $X^{\text{ext}}$ and $Y^{\text{ext}}$, doing so based on the \textit{independence of cause and mechanism} postulate \citep{Daniusis2010}
\begin{postulate}\label{postulate1}
The mechanisms generating the random variable describing the cause, denoted by $X$, and generating the random variable describing the effect given the cause, denoted by $Y\mid X$, are independent, i.e., they contain no information about each other.
\end{postulate}
The independence of mechanisms is related to Pearl's notion of stability \cite[Section 2.4]{Pearl2000} which states that the causal mechanism describing a variable given its cause must remain unchanged if one changes the mechanism generating the cause. For instance, consider the toy example where the altitude of a station modelled by a random variable $X$ and the temperature at this station modelled by the random variable $Y$ are causally related through the structural equation model \citep{Pearl2000,pearl2016causal} 
\begin{equation}
Y= h(X, \epsilon), \quad X \independent \epsilon, \notag
\end{equation}
where $h$ is the mechanism (function) describing $Y \mid X$ and which can be thought of as modelling the physical mechanism relating the temperature to the altitude, and $\epsilon$ is an error term. Then, any localised intervention on the altitude, e.g., by considering a nearby station, would result in a change in the temperature but would not affect the physical mechanism $h$. Therefore, the conditional random variable $Y \mid X$ described by $h$ provides no information about $X$.

\cite{Janzing2010} formalize the notion of two mechanisms containing no information about each other using algorithmic information theory and more specifically the notion of Kolmogorov complexity, that is \textit{the length of the shortest computer program that prints a sequence of the underlying random variable and halts \citep{ Kolmogorov1968,Li2008}}. The Kolmogorov complexity of a random sequence is closely related to the Shannon entropy \citep{Shannon1948} of the underlying distribution. For a random sequence $X^n= \{ X_i \}_{i=1}^n$, with $X_i$ drawn independently from a probability distribution $F$ with density $f$, \cite{Brudno1983} shows that the Kolmogorov complexity $K(X^n)$ of the sequence $X^n$ is linked to the Shannon entropy of $F$ through
\begin{equation}
 \underset{n \rightarrow \infty}{\lim} \dfrac{K(X^n)}{n} = - \int f(x) \log_2 f(x) dx, \quad a.s. \label{KC_Shannon}
\end{equation}
This result is also valid for discrete distributions where the Shannon entropy in the right hand-side of \eqref{KC_Shannon} is modified accordingly. For example, if $X^n \in \{0,1\}^n$ is drawn from $n$ independent Bernoulli random variables with known success probability $p$, then, with probability close to 1, $K(X^n)$ is close to $n$ times the binary entropy, i.e., $-np \log_2(p) - n(1-p) \log_2(1-p)$.

In terms of the Kolmogorov complexities of the random variables $X$ and $Y$, Postulate~\ref{postulate1} is translated as follows:
\begin{postulate}\label{postulate2}
If the random variable $X$ is the cause of a random variable $Y$, then the distribution of $X$, $F_X$, and the distribution of $Y\mid X$, $F_{Y \mid X}$, are algorithmically independent, that is 
$$K(F_{X,Y}) = K(F_X)+K(F_{Y\mid X}) $$
where $K(F_Z)$ stands for the Kolmogorov complexity of a random variable $Z \sim F_Z$.
\end{postulate}
In a setting where $X$ and $Y$ are causally related, i.e., either $X$ causes $Y$ or $Y$ causes $X$, Postulate~\ref{postulate2} implies that one should infer that $X$ causally influences $Y$ whenever
\begin{equation}
K(F_X)+K(F_{Y\mid X}) \overset{+}{\leq} K(F_Y) + K(F_{X\mid Y}),\label{algo_indep}
\end{equation}
where $\overset{+}{\leq}$ denotes inequality up to an additive constant. This inequality stems from the definition of the algorithmic mutual information of two random variables $U$ and $V$, which is a positive quantity equal to $I(F_U:F_V)= K(F_V)- K(F_{U,V})+K(F_U)$ and from the equivalence between $K(F_{U,V})$ and $K(F_{U,V\mid U})$; see \citet[Section II-A]{Janzing2010} for details. One possible setting where $X$ and $Y$ are not causally related is where $X$ and $Y$ are not causally sufficient, i.e., there is a latent common cause or unobserved confounder to $X$ and $Y$. In the latter case, the joint distribution function $F_{X,Y}$ might have a smaller complexity $K(F_{X,Y})$ when considering its decomposition by the chain rule involving the marginal distribution of the confounder. We assume in this work the absence of common confounders and proceed with the weaker version of Postulate~\ref{postulate2} given by \eqref{algo_indep}.

\subsection{\textsf{Kolmogorov complexity and code length}\label{subsec:KC}}
Here we describe the link between the Kolmogorov complexity of a random variable and its code length through the minimum description length (MDL) of \cite{RISSANEN1978}. The MDL translates Occam's razor, i.e., the parsimony principle that complex models should not be used beyond necessity, using information theory and states that one should choose the model that provides the shortest description of the data.
%The idea of describing an object goes back to the descriptive complexity theory of Kolmogorov and its use for expressing random events through the length of the shortest computer program that prints a sequence of the underlying random variable and terminates, known as the Kolmogorov complexity. 
Random objects being incompressible, they cannot have a concise description \citep{Turing1937} and their descriptive Kolmogorov complexity is therefore not computable \citep{Cover1991}. Rissanen modified the concept of the Kolmogorov complexity by proposing the MDL, which focuses on the description length of probability distributions. This paves the way to new insights into many statistical procedures \citep{Hansen2001c,Davis2006, Aue2014}. From Rissanen's perspective and based on Shannon's source coding theorem \citep{Shannon1948}, the MDL of the data is the descriptive power, also called the code length, of its underlying generating distribution. In the example of the random sequence $X^n \in \{0,1\}^n$ described above, one can encode every symbol of the sequence at a cost of $-\log_2(p)$ for a $1$ and $-\log_2(1-p)$ for a $0$, resulting in a total code length of $X^n$ equal to the negative log-likelihood of the Bernoulli model. We denote this quantity by $CL_{\mathcal{M}}(X^n)$, where in this example $\mathcal{M}$ describes the Bernoulli model with known probability of success $p$.

As the true distribution of the data is rarely known, we minimize the code length of many probability distributions from a specific model class with the same sample space as the data. Based on the MDL principle, the Kolmogorov complexity of a random variable can be practically approximated by its code length, which is defined relative to a model class (whether it contains the true underlying distribution or not) and computed using a specific coding scheme. This approximation allows us to formulate Postulate~\ref{postulate2} in terms of code lengths leading to the following inequality whenever a random variable $X$ is the cause of a random variable $Y$:
\begin{equation}
CL_{\mathcal{M}_X}(X) + CL_{\mathcal{M}_{Y\mid X}}(Y\mid X) \leq CL_{\mathcal{M}_{Y}}(Y) + CL_{\mathcal{M}_{X\mid Y}}(X\mid Y), \label{causal_CL}
\end{equation}
where each quantity is to be understood as dependent on the observed dataset.
%where $CL_{\mathcal{M}_Z}(Z)$ denotes the code length of a random variable $Z$ under model class $\mathcal{M}_Z$.
\subsection{\textsf{Causal discovery using quantile scoring}}
The MDL principle defines the ``best" fitting model from a model class as the one within this class that produces the shortest code length that completely describes the observations. Different forms of the model-based code length have been devised using various coding algorithms \citep[Section 3]{Hansen2001c} but we focus on the ``two-stage" or ``two-part" version, stating that the code length of an observed sequence of a random variable can be decomposed into a sum of two parts. The first represents the code length of the fitted model, that is the amount of space required to store the fitted model or, in a fully parametric setting, the corresponding estimated parameter. The second part of the MDL represents the code length of the data based on the fitted model and is equal to the negative log-likelihood of the model evaluated at the transmitted fitted parameter, i.e., the negative log-likelihood of the fitted model. In the example of the Bernoulli sequence $X^n$, as the probability of success $p$ is known, we showed in Section~\ref{subsec:KC} that the code length of the sequence is equal to the negative log-likelihood of the Bernoulli model, i.e., it is equal to the second part of the MDL. When the probability of success $p$ is unknown, the two-stage MDL takes the form of a penalised log-likelihood where the penalty, computed in its first part, is the complexity of estimating $p$.

We derive the code lengths of the marginal and conditional random variables in the dataset $\{(X^{\text{ext}}_i,Y_i^{\text{ext}})\}_{i=1}^{n_u}$ using a quantile-based MDL obtained with respect to a specific model class. That is, we encode the dataset of the tail observations using their quantile function under the chosen model class rather than their distribution function. We use a quantile-based approach as it can provide an enhanced characterization of the distribution of an outcome variable by capturing features such as heteroskedasticity that mean-based approaches such as that in \cite{Peters2016} will fail to capture. In contrast with \cite{Tagasovska2018} who focus on causal discovery in the main body of the observations using the MDL principle in a nonparametric setting, we concentrate on causal discovery in the tail region relying on the MDL principle within the class of extreme value distributions.

As described in Section~\ref{sec:joint_exc}, the distribution function of the joint tails of the vector $(X,Y)$ can be approximated by a bivariate distribution with shifted GPD margins and an extreme value copula $C^{EV} \in \mathcal{C}^{EV}$ satisfying~\eqref{ev_pickands} for some valid Pickands dependence function $A$. More specifically, denoting the marginal distributions $F= F_{(\sigma_X,\xi_X)}\sim \text{GPD}(u_X,\sigma_X,\xi_X)$ and $G= G_{(\sigma_Y,\xi_Y)}\sim \text{GPD}(u_Y,\sigma_Y,\xi_Y)$, we have
\begin{eqnarray}
\Pr(X^{\text{ext}} \leq x, Y^{\text{ext}} \leq y) &=& C^{EV} \{F(x), G(y)\}, \quad x>u_X, y>u_Y, \notag \\
\Pr({X^\text{ext}} \leq x \mid Y^{\text{ext}} = y>u_Y) &=& \partial_{v_2} C^{EV} \{F(x), G(y)\}, \quad x>u_X, \notag \\
\Pr(Y^{\text{ext}} \leq y \mid X^{\text{ext}} = x>u_X) &=& \partial_{v_1} C^{EV} \{F(x), G(y)\}, \quad y>u_Y. \notag
\end{eqnarray}
We consider a marginal model for the $\tau$-th quantiles of $X^{\text{ext}}$ and $Y^{\text{ext}}$,
\begin{eqnarray}
Q_{X^{\text{ext}}}(\tau) &=& F^{-1}(\tau),  \label{marginal_model_X} \\
Q_{Y^{\text{ext}}}(\tau) &=& G^{-1}(\tau),  \label{marginal_model_Y} 
\end{eqnarray}
as well as the following model for the conditional $\tau$-th quantiles of $X^{\text{ext}}\mid Y^{\text{ext}}$ and $Y^{\text{ext}} \mid X^{\text{ext}}$
 \begin{eqnarray}
Q_{X^{\text{ext}}\mid Y^{\text{ext}}=y>u_Y}(\tau) &=& F^{-1} \left[  (\partial_{v_2} C^{EV})^{-1} \left\lbrace  \tau,G(y) \right\rbrace  \right] , \label{conditional_model_X_Y} \\
Q_{Y^{\text{ext}} \mid X^{\text{ext}}=x>u_X}(\tau) &=& G^{-1} \left[ (\partial_{v_1} C^{EV})^{-1} \left\lbrace F(x),\tau \right\rbrace  \right] ,  \label{conditional_model_Y_X}
\end{eqnarray}
where
\begin{eqnarray}
(\partial_{v_2} C^{EV})^{-1} \{\tau,G(y)\}  &=& \text{inf} \left[  v \in [0,1] : \partial_{v_2} C^{EV} \{ v,G(y) \} = \tau \right], \notag \\
(\partial_{v_1} C^{EV})^{-1} \{F(x),\tau\} &=& \text{inf} \left[  v \in [0,1] : \partial_{v_1} C^{EV} \{ F(x),v \} = \tau \right] . \notag
\end{eqnarray}
Here, the level $\tau$ ranges from a probability of $0$ to a probability of $1$ covering the entire joint tail distribution. We denote the corresponding classes of models by $\mathcal{M}_{X^{\text{ext}}}$, $\mathcal{M}_{Y^{\text{ext}}}$, $\mathcal{M}_{X^{\text{ext}}\mid Y^{\text{ext}}}$, and $\mathcal{M}_{Y^{\text{ext}} \mid X^{\text{ext}}}$ and any model from these classes by $\mathcal{F}_{X^{\text{ext}}} \in \mathcal{M}_{X^{\text{ext}}}$, $\mathcal{F}_{Y^{\text{ext}}} \in \mathcal{M}_{Y^{\text{ext}}}$, $\mathcal{F}_{X^{\text{ext}} \mid Y^{\text{ext}}} \in \mathcal{M}_{X^{\text{ext}} \mid Y^{\text{ext}}}$, and $\mathcal{F}_{Y^{\text{ext}} \mid X^{\text{ext}}} \in \mathcal{M}_{Y^{\text{ext}} \mid X^{\text{ext}}}$. The superscripts $\text{ext}$ are omitted where no confusion can arise. If $CL_{\mathcal{F}_X}^{\tau}(X^{\text{ext}})$ denotes the code length of the random variable $X^{\text{ext}}$ under the $\tau$-th quantile model $\mathcal{F}_X \in \mathcal{M}_{X}$, then
\begin{equation}
CL_{\mathcal{F}_X}^{\tau}(X^{\text{ext}}) = CL_{\mathcal{F}_X}^{\tau}(\hat{\mathcal{F}}_X) + CL_{\mathcal{F}_X}^{\tau}(\hat{\mathcal{\epsilon}}_X \mid \hat{\mathcal{F}}_X), \notag
\end{equation}
where $CL_{\mathcal{F}_X}^{\tau}(\hat{\mathcal{F}}_X)$ is the code length of the fitted model $\hat{\mathcal{F}}_X$ and $CL_{\mathcal{F}_X}^{\tau}(\hat{\mathcal{\epsilon}}_X \vert \hat{\mathcal{F}}_X)$ is the leftover information not captured by the transmitted model $\hat{\mathcal{F}}_X$, i.e., the code length of the residuals of $\hat{\mathcal{F}}_X$; see \cite{Davis2006,Aue2014} for related approaches in the settings of autoregressive and quantile regression modelling, respectively. We first focus on encoding the first part of the ``two-stage" version of the MDL of $X^{\text{ext}}$ and $Y^{\text{ext}}$. Since any $\hat{\mathcal{F}}_X$ or $\hat{\mathcal{F}}_Y$ is completely specified by the scale and shape parameters of the GPD, the code lengths for encoding the fitted models $\hat{\mathcal{F}}_X$ and $\hat{\mathcal{F}}_Y$ are equal to those of the estimates of the parameters $\sigma_X$ and $\xi_X$ and $\sigma_Y$ and $\xi_Y$. The fitting step is performed using maximum likelihood (ML) estimation and we can use the result by \cite{Rissanen1989} stating that the code length of a ML estimate of a real-valued parameter based on $n$ observations is equal to $\log_2(n)/2$, to obtain
\begin{equation}
CL_{\mathcal{F}_X}^{\tau}(\hat{\mathcal{F}}_X) = CL_{\mathcal{F}_Y}^{\tau}(\hat{\mathcal{F}}_Y)= \dfrac{p}{2} \log_2(n_u), \notag
\end{equation}
where $p=2$ is the number of parameters of the GPD. The fitted conditional models $\hat{\mathcal{F}}_{X\mid Y}$ and $\hat{\mathcal{F}}_{X\mid Y}$ are more complicated to encode, as the estimation procedure involves both the estimation of the margins, using ML, and the estimation of the dependence structure described by the extreme value copula $C^{EV}$, which is estimated non-parametrically as described in Section~\ref{sec:inference_ev_copula}. As the fitted conditional models rely on the same marginal distributions and non-parametric extreme value copula, their code lengths $CL^{\tau}_{\mathcal{F}_{X\mid Y}}(\hat{\mathcal{F}}_{X\vert Y})$ and $CL^{\tau}_{\mathcal{F}_{Y\mid X}}(\hat{\mathcal{F}}_{Y\mid X})$ are equal, and we show below that an analytical expression of their complexity is not needed for causal discovery between $X^{\text{ext}}$ and $Y^{\text{ext}}$. 

We now encode the residuals $\hat{\mathcal{\epsilon}}_X$, $\hat{\mathcal{\epsilon}}_Y$, $\hat{\mathcal{\epsilon}}_{X\mid Y}$, and $\hat{\mathcal{\epsilon}}_{Y\mid X}$ of the fitted models. That is, we compute the second part of the ``two-stage" MDL for the $\tau$-th quantile models. As discussed above, this second part is equal to the negative log-likelihood of the model evaluated at the fitted parameters. We derive such quantities relying on the link between the asymmetric Laplace density and quantile regression \citep{Komunjer2005}. More precisely, following \cite{Geraci2007} and \cite{Aue2014}, the code lengths of the
innovations in our $\tau$-th quantile models \eqref{marginal_model_X}--\eqref{conditional_model_Y_X} are obtained through an application of the asymmetric Laplace likelihood functions
\begin{eqnarray}
L(\hat{\sigma}_X,\hat{\xi}_X) &=& \tau^{n_u} (1-\tau)^{n_u} \exp\{-\hat{S}_{X^{\text{ext}}}(\tau)\}, \notag \\
L(\hat{\sigma}_Y,\hat{\xi}_Y) &=& \tau^{n_u} (1-\tau)^{n_u} \exp\{-\hat{S}_{Y^{\text{ext}}}(\tau)\}, \notag \\
L(\hat{\sigma}_X,\hat{\sigma}_Y,\hat{\xi}_X,\hat{\xi}_Y,\hat{C}^{EV}) &=& \tau^{n_u} (1-\tau)^{n_u} \exp\{-\hat{S}_{X^{\text{ext}}\mid Y^{\text{ext}}}(\tau)\}, \notag \\
L(\hat{\sigma}_X,\hat{\sigma}_Y,\hat{\xi}_X,\hat{\xi}_Y,\hat{C}^{EV}) &=& \tau^{n_u} (1-\tau)^{n_u} \exp\{-\hat{S}_{Y^{\text{ext}}\mid X^{\text{ext}}}(\tau)\}, \notag
\end{eqnarray}
respectively. The quantities $\hat{S}_{X^{\text{ext}}}(\tau)$, $\hat{S}_{Y^{\text{ext}}}(\tau)$, $\hat{S}_{X^{\text{ext}}\mid Y^{\text{ext}}}(\tau)$, and $\hat{S}_{Y^{\text{ext}} \mid X^{\text{ext}}}(\tau)$ are the estimates of the expected quantile scores (multiplied by $n_u$) of the $\tau$-th quantile forecasts \eqref{marginal_model_X}--\eqref{conditional_model_Y_X} \citep{Koenker1999,Gneiting2007}, and are given by
%\begin{scriptsize}
%\begin{adjustwidth*}{0cm}{-2.7cm}
\begin{eqnarray}
\hat{S}_{X^{\text{ext}}}(\tau) &=&  \sum_{i=1}^{n_u} \rho_{\tau} \left\lbrace X^{\text{ext}}_{i} - F^{-1}_{(\hat{\sigma}_X,\hat{\xi}_X)} (\tau)\right\rbrace , \notag \\
\hat{S}_{Y^{\text{ext}}}(\tau) &=&\sum_{i=1}^{n_u} \rho_{\tau} \left\lbrace Y^{\text{ext}}_{\i} - G^{-1}_{(\hat{\sigma}_Y,\hat{\xi}_Y)} (\tau)\right\rbrace, \notag \\
\hat{S}_{X^{\text{ext}} \mid Y^{\text{ext}}}(\tau) &=& \sum_{i=1}^{n_u} \rho_{\tau} \left( X^{\text{ext}}_{i} - F_{(\hat{\sigma}_X,\hat{\xi}_X)}^{-1}\left[  (\partial_{v_2} \hat{C}^{EV})^{-1} \{\tau,G_{(\hat{\sigma}_Y,\hat{\xi}_Y)}(Y^{\text{ext}}_{i})\} \right] \right) , \notag \\
\hat{S}_{Y^{\text{ext}} \mid X^{\text{ext}}}(\tau) &=& \sum_{i=1}^{n_u} \rho_{\tau} \left(  Y^{\text{ext}}_{i} - G_{(\hat{\sigma}_Y,\hat{\xi}_Y)}^{-1}\left[  (\partial_{v_1} \hat{C}^{EV})^{-1} \{F_{(\hat{\sigma}_X,\hat{\xi}_X)}(X^{\text{ext}}_{i}),\tau\} \right] \right), \notag
\end{eqnarray}
where $\rho_{\tau}$ is a loss function given by the check function $\rho_{\tau}(t)= (\mathbf{1}_{\{t \geq 0\}}-\tau)t$ \citep{Koenker1999}, and $(\hat{\sigma}_X,\hat{\sigma}_Y,\hat{\xi}_X,\hat{\xi}_Y) \in \mathbb{R}^2_{+}\times \mathbb{R}^2$ and $\hat{C}^{EV} \in \mathcal{C}^{EV}$ are the transmitted fitted parameters.

\bigskip
%Even though the innovations do not necessarily follow an asymmetric
%Laplace distribution, inference procedures are based on the AL density
%but in a quasi-likelihood setting and 
%The parameter estimates 
%of the specified marginal and conditional models 
%\eqref{marginal_model_X}--\eqref{conditional_model_Y_X} are
%consistent \citep{Komunjer2005}. 
Thus, summing both parts of the MDL,
the code lengths of our random variables are
\begin{eqnarray}
CL^{\tau}_{\mathcal{F}_{X}}(X^{\text{ext}}) &=& \log_2(n_u) + \hat{S}_{X^{\text{ext}}}(\tau) - n_u \log\{\tau (1-\tau)\}, \notag \\
CL^{\tau}_{\mathcal{F}_{Y}}(Y^{\text{ext}}) &=& \log_2(n_u) + \hat{S}_{Y^{\text{ext}}}(\tau) - n_u \log\{\tau (1-\tau)\}, \notag \\
CL^{\tau}_{\mathcal{F}_{X\mid Y}}(X^{\text{ext}}\mid Y^{\text{ext}}) &=& CL^{\tau}_{\mathcal{F}_{X\mid Y}}(\hat{\mathcal{F}}_{X\vert Y}) + \hat{S}_{X^{\text{ext}}\mid Y^{\text{ext}}}(\tau) - n_u \log\{\tau (1-\tau)\}, \notag \\
CL^{\tau}_{\mathcal{F}_{Y\mid X}}(Y^{\text{ext}}\mid X^{\text{ext}}) &=& CL^{\tau}_{\mathcal{F}_{Y\mid X}}(\hat{\mathcal{F}}_{Y\mid X}) + \hat{S}_{Y^{\text{ext}}\mid X^{\text{ext}}}(\tau) - n_u \log\{\tau (1-\tau)\}. \notag
\end{eqnarray}

Finally, as the complexities of the fitted conditional models are equal, we can formulate Postulate~\ref{postulate2} in terms of the $\tau$-th quantile scores through the equivalence between the inequalities
\begin{eqnarray}
CL_{\mathcal{M}_X}(X^{\text{ext}}) + CL_{\mathcal{M}_{Y\mid X}}(Y^{\text{ext}}\mid X^{\text{ext}}) &\leq& CL_{\mathcal{M}_{Y}}(Y^{\text{ext}}) + CL_{\mathcal{M}_{X\mid Y}}(X^{\text{ext}}\vert Y^{\text{ext}}) \notag \\
&\Leftrightarrow \notag \\
\hat{S}_{X^{\text{ext}}}(\tau) + \hat{S}_{Y^{\text{ext}}\mid X^{\text{ext}}}(\tau) &\leq& \hat{S}_{Y^{\text{ext}}}(\tau) + \hat{S}_{X^{\text{ext}}\mid Y^{\text{ext}}}(\tau). \label{causal_QS}
\end{eqnarray}

The causality model at extreme levels is expected to be stable with respect to the quantiles, i.e., the inequality~\eqref{causal_QS} is expected to hold for various levels $\tau$. By defining $\hat{S}_{X^{\text{ext}}}= \int_{0}^1 \hat{S}_{X^{\text{ext}}}(\tau)d \tau$ (a similar notation holds for $Y^{\text{ext}}$, $X^{\text{ext}}\mid Y^{\text{ext}}$, and $Y^{\text{ext}}\mid X^{\text{ext}}$), we modify our decision rule about causality from \eqref{causal_QS} to 
\begin{equation}
\hat{S}_{X^{\text{ext}}} + \hat{S}_{Y^{\text{ext}}\mid X^{\text{ext}}} \leq \hat{S}_{Y^{\text{ext}}} + \hat{S}_{X^{\text{ext}}\mid Y^{\text{ext}}}. \notag
\end{equation}
Further, we define the causal score of our CausEV method as 
\begin{equation}
S_{X \rightarrow Y}^{\text{ext}} = \dfrac{\hat{S}_{Y^{\text{ext}}} + \hat{S}_{X^{\text{ext}}\mid Y^{\text{ext}}}}{\hat{S}_{X^{\text{ext}}} + \hat{S}_{Y^{\text{ext}}\mid X^{\text{ext}}}+\hat{S}_{Y^{\text{ext}}} + \hat{S}_{X^{\text{ext}}\mid Y^{\text{ext}}}} , \label{QCDD_extremes}
\end{equation}
and conclude that $X$ causes $Y$ at extreme levels whenever $S_{X \rightarrow Y}^{\text{ext}}> 0.5$, where equality stands for the non-identifiable setting \citep{Peters2014}, i.e., a setting where one cannot identify the causal direction at extreme levels between $X$ and $Y$.

%Further, we conclude that $X$ causes $Y$ at extreme levels whenever
%\begin{equation}
%S_{X \rightarrow Y}^{\text{ext}} = \dfrac{\hat{S}_{Y^{\text{ext}}} + \hat{S}_{X^{\text{ext}}\mid Y^{\text{ext}}}}{\hat{S}_{X^{\text{ext}}} + \hat{S}_{Y^{\text{ext}}\mid X^{\text{ext}}}+\hat{S}_{Y^{\text{ext}}} + \hat{S}_{X^{\text{ext}}\mid Y^{\text{ext}}}} > 0.5, \label{QCDD_extremes}
%\end{equation}
%where equality stands for the non-identifiable setting \citep{Peters2014}, i.e., a setting where one cannot identify the causal direction at extreme levels between $X$ and $Y$.
%%% the code length of the fitted model of GPD(\sigma,\xi) = code length of GPD(\tilde{\sigma},\tilde{\xi})

\section{\textsf{Simulation study}}
\label{simul}
We show the effectiveness of CausEV under different simulation scenarios
and compare it to state-of-the-art methods for uncovering causality at the mean
level, such as LINGAM \citep{Shimizu2006}, IGCI \citep{Janzing2012}, CAM \citep{Buhlmann2014}, and RESIT \citep{Peters2014}.

\subsection{\textsf{Additive noise models}}
We consider a structural equation model with an additive noise (AN) 
structure between $X$ and $Y$, i.e., 
\begin{equation}
Y = h(X) + \epsilon, \quad X \independent \epsilon, \label{AN_structure}
\end{equation}
where the deterministic structural function $h$ and the 
distribution functions of $X$ and $\epsilon$, denoted $F_X$ and
$F_{\epsilon}$ respectively, are such that the AN structure
\eqref{AN_structure} holds in the joint upper tail of $(X,Y)$.
This avoids scenarios where large values of $X$ 
induce low values for $Y$. Moreover, we require asymptotic dependence between $X$ and $Y$, i.e., the 
joint distribution of $(X,Y)$ must be in the maximum domain of 
attraction of some dependent MEVD; see Section~\ref{sec:MEVT}. Under the AN structure \eqref{AN_structure}, the coefficient of tail dependence $\chi$ defined in \eqref{chi_formula}, is equal to
\begin{eqnarray}
\chi_u &=& \dfrac{1}{u} \int_{F_X^{-1}(u)}^{+\infty} \left[ 1- F_{\epsilon} \left\lbrace F_Y^{-1}(u) - h(x) \right\rbrace \right] f_{X}(x) dx,\notag \\
F_Y^{-1}(u) &=& \text{inf} \left\lbrace y \in \mathbb{R} : \int_{-\infty}^{y} \int_{-\infty}^{+\infty} f_{\epsilon}(t-s) f_{h(X)} (s) ds dt \geq u \right\rbrace . \notag
\end{eqnarray}
In the absence of the noise random variable $\epsilon$, a monotonic increasing function $h$ ensures comonotonicity between $X$ and $Y$ and hence $\chi=1$. We monitor the effect of the noise variable that should result in $\chi <1$ while maintaining asymptotic dependence. We consider the following settings for the AN structure \eqref{AN_structure}:
\begin{description}
\item[Scenario 1.] $X \sim \text{GPD}(2,0.3,0.1)$
\begin{enumerate}
\item[(a)] $h(x)= \log(x+10)+x^6$ and $\epsilon \sim t(\nu_{\epsilon})$, with $\nu_{\epsilon} \in [2.1,4]$,
\item[(b)] $h(x)= x^3+x$ and $\epsilon \sim \mathcal{N}(0,\sigma_{\epsilon}^2)$, with $\sigma_{\epsilon} \in [0.1,20]$;
\end{enumerate}
\item[Scenario 2.] $X \sim \mathcal{N}(1,0.4^2)$
\begin{enumerate}
\item[(c)] $h(x)= \log(x+10)+x^6$ and $\epsilon \sim t(\nu_{\epsilon})$, with $\nu_{\epsilon} \in [2.1,4]$,
\item[(d)] $h(x)= x^3+x$ and $\epsilon \sim \mathcal{N}(0,\sigma_{\epsilon}^2)$, with $\sigma_{\epsilon} \in [0.05,4]$;
\end{enumerate}
\item[Scenario 3.] $X \sim \text{GEV}(-2.8,1,-0.1)$
\begin{enumerate}
\item[(e)] $h(x)= x^3+x$ and $\epsilon \sim \mathcal{N}(0,\sigma_{\epsilon}^2)$, with $\sigma_{\epsilon} \in [0.05,4]$.
\end{enumerate}
\end{description}

\begin{figure}[!h]
	\centering 
 \subfloat{\includegraphics[width=0.33\textwidth]{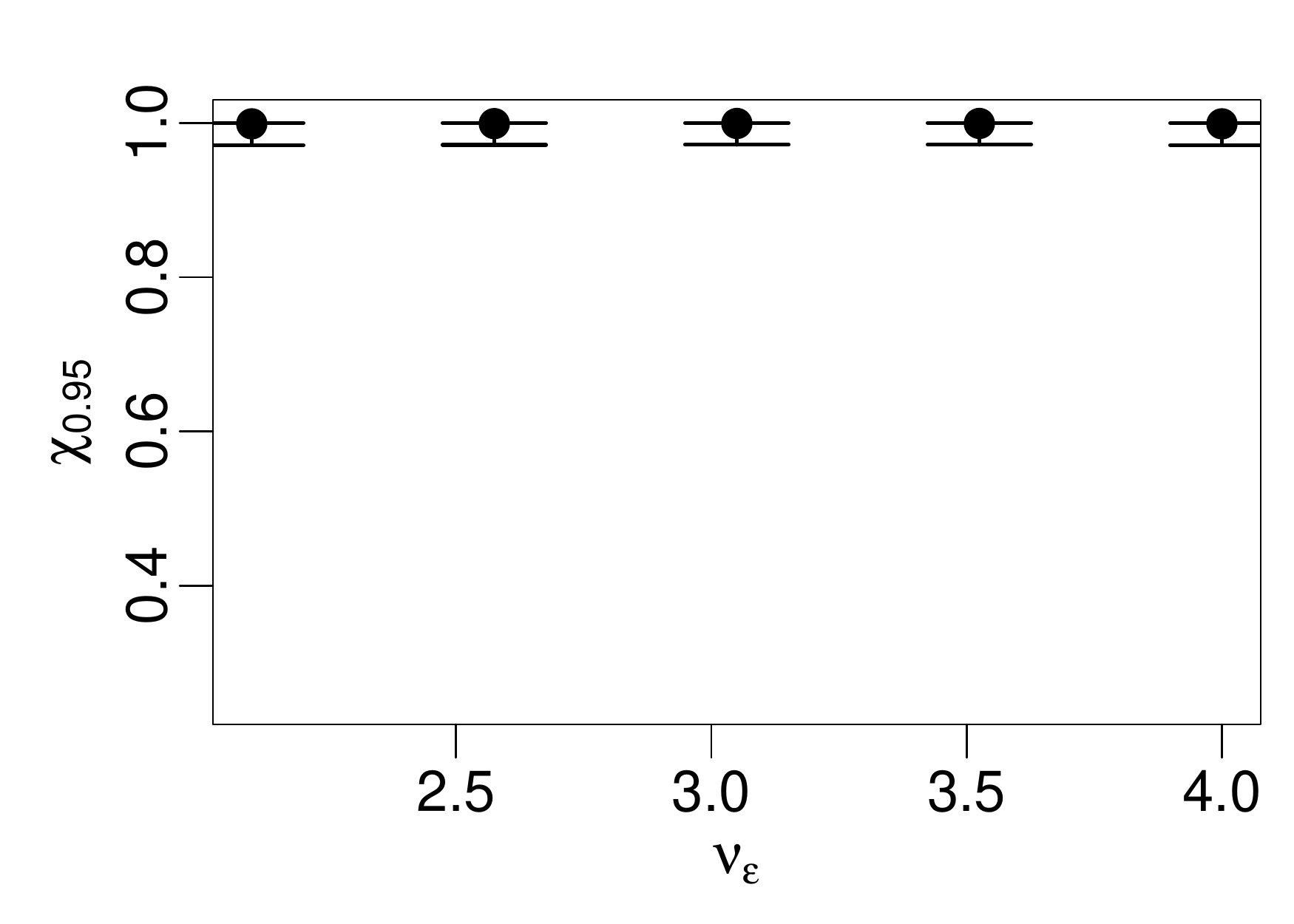}} \hfill
 \subfloat{\includegraphics[width=0.33\textwidth]{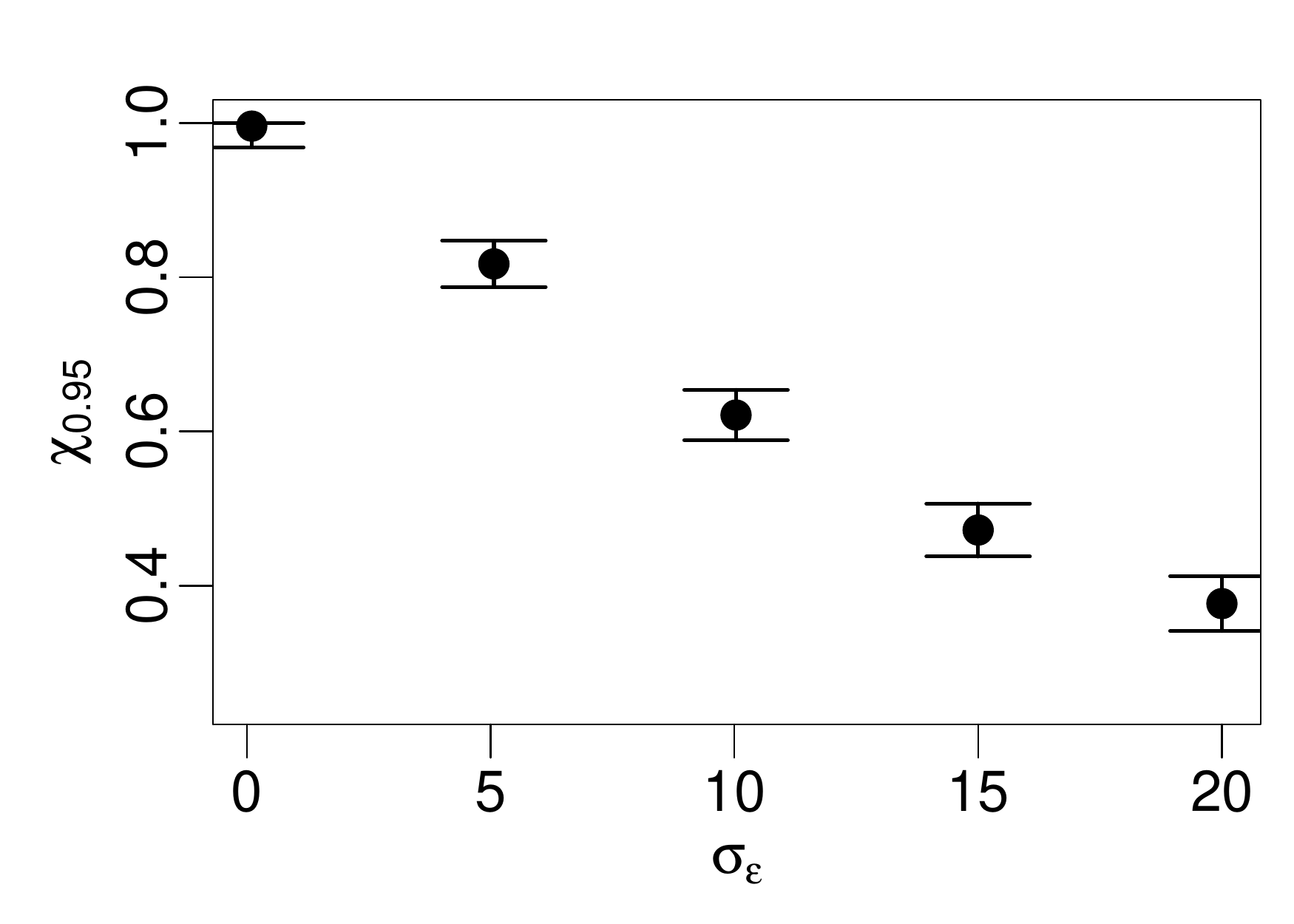}} \hfill
  \subfloat{\includegraphics[width=0.33\textwidth]{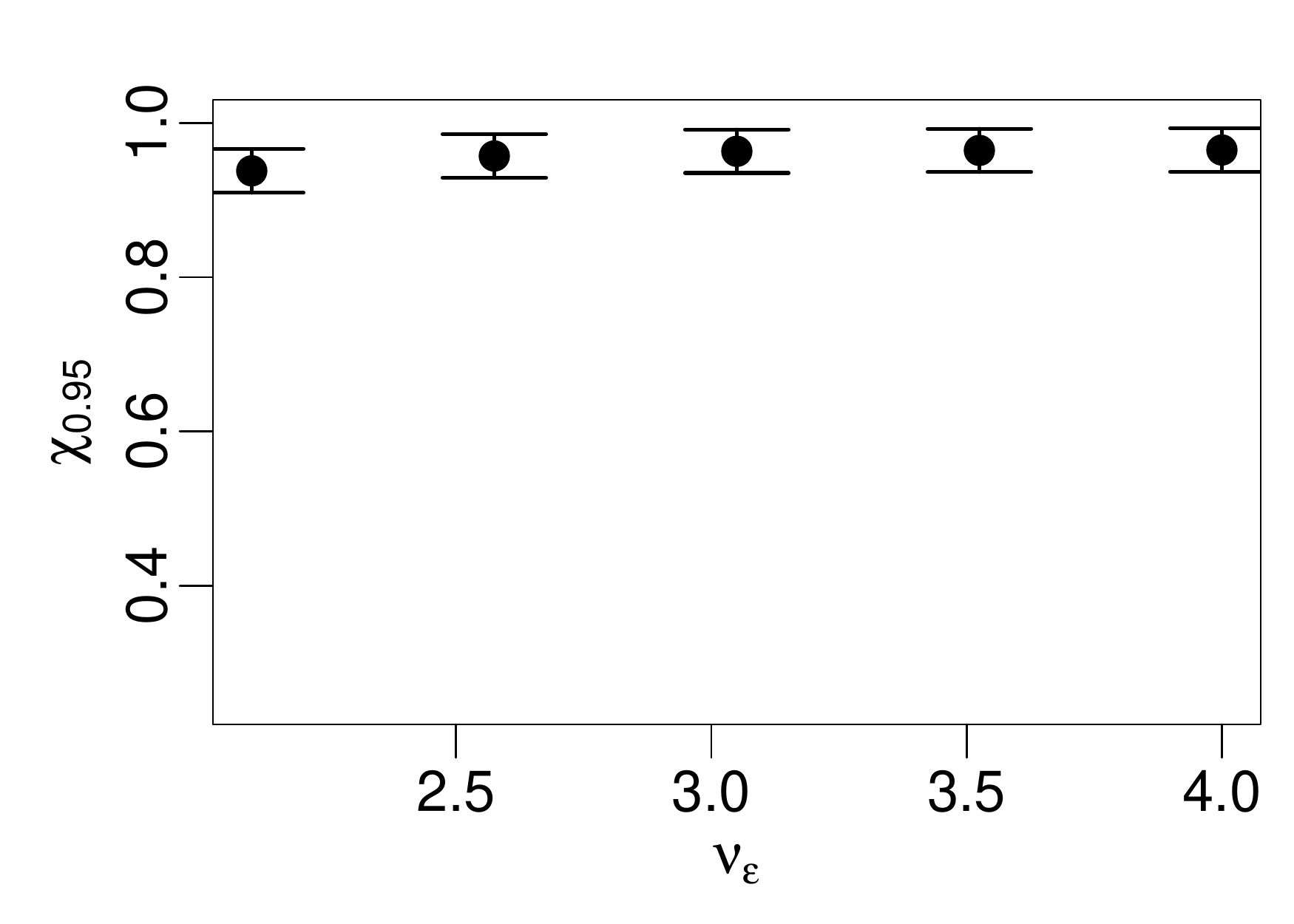}} \\
   \subfloat{\includegraphics[width=0.33\textwidth]{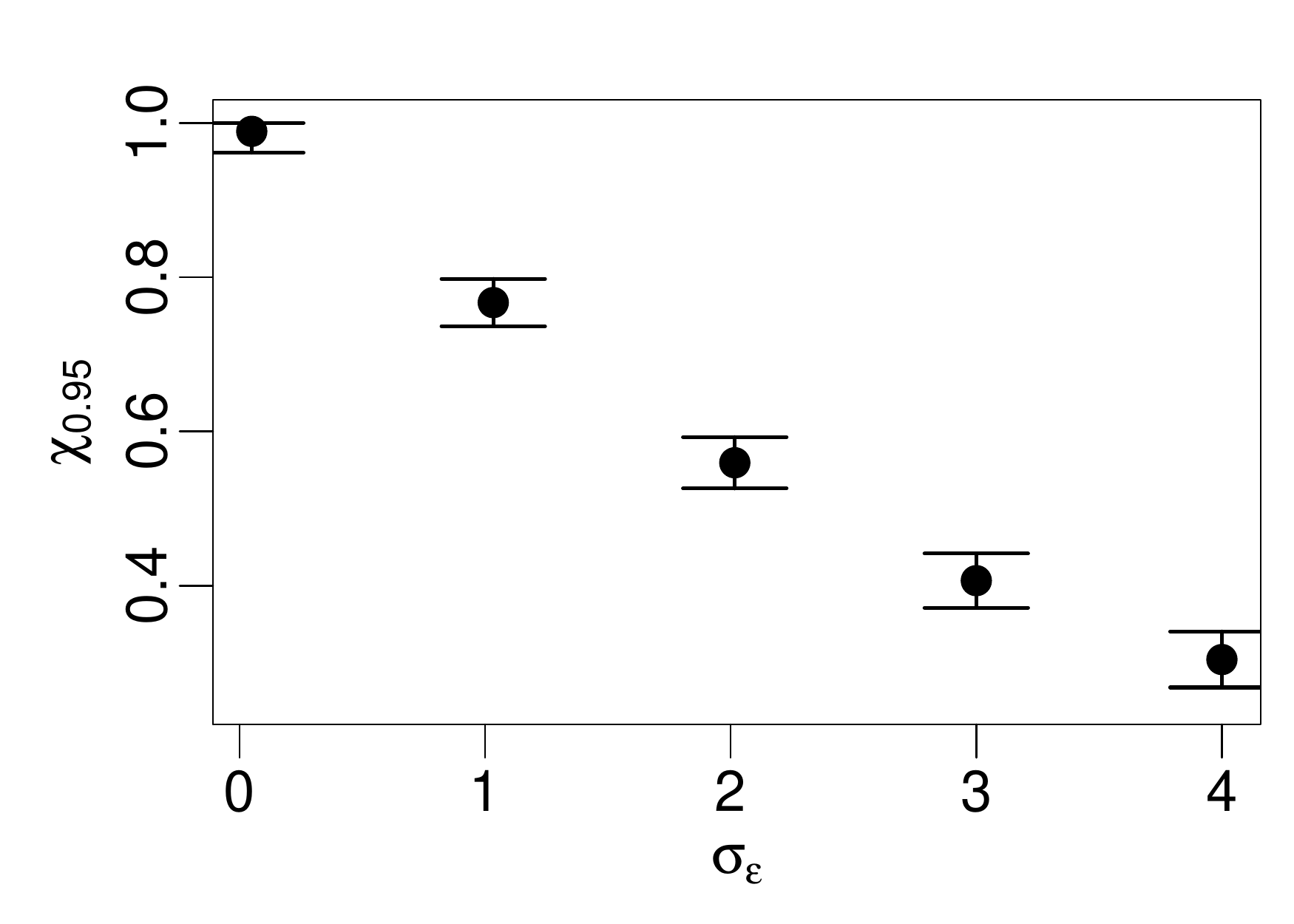}} \hfill
  \subfloat{\includegraphics[width=0.33\textwidth]{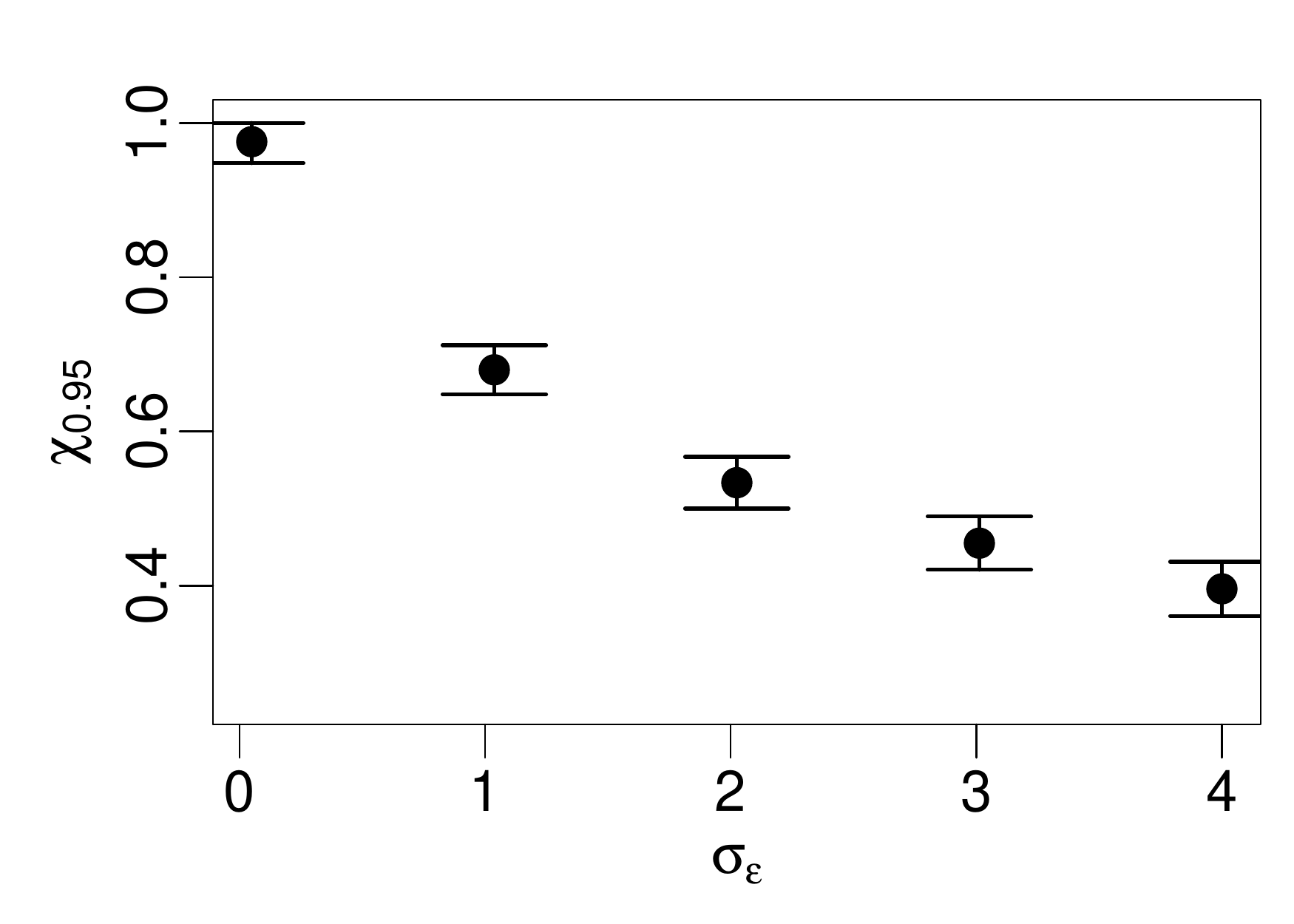}}
	\caption{Empirical estimates of $\chi_{0.95}$ for scenarios 1(a) (upper left panel); 1(b) (upper middle panel); 2(c) (upper right panel); 2(d) (lower left panel) and 3(e) (lower right panel). The $x$-axis represents either the degrees of freedom $\nu_{\epsilon}$ or the standard deviation $\sigma_{\epsilon}$.}
	\label{fig:chi_anm}
\end{figure}

Figure~\ref{fig:chi_anm} displays empirical estimates of $\chi_{0.95}$ 
computed under Scenarios 1--3. The resulting tail dependence coefficients
hint at asymptotic dependence of $(X,Y)$ in all cases. We proceed 
with CausEV to distinguish the cause ($X$) from the
effect ($Y$) in the upper quadrant
$\{(X^{\text{ext}}_i,Y^{\text{ext}}_i)\}_{i=1}^{n_u} = 
\{ (X_i,Y_i) : X_i>F^{-1}_X(0.95) \ \text{and } Y_i>F^{-1}_Y(0.95)\}$.
Experiments are based on $300$ repetitions where we fix the size of 
concurrent exceedances at $n_u=55$. 

%\subsubsection*{\textsf{Performance of CausEV}}
Owing to the relatively small size of the set of extreme observations,
we perform our score-based CausEV relying solely on three quantiles
of the limiting GPD given by Legendre quadrature of the
interval $[0,1]$, i.e., the quantities in the relation~\eqref{causal_QS}
are computed for $\tau=0.5$ and $\tau=0.5 \times (1\pm \sqrt{3/5})$ and 
weighted respectively by $w=0.5 \times (5/9)$ and $w=0.5 \times (8/9)$ to
approximate the integrated scores in \eqref{QCDD_extremes}. We conduct
the estimation of the extreme value copula using the min-projection method of \cite{Mhalla_OT_CD} at $500$ unequally spaced values in the unit simplex.
The margins are transformed to the unit Fr\'{e}chet scale using
rank transformations.

\begin{figure}[!h]
	\centering 
 \subfloat{\includegraphics[width=0.33\textwidth]{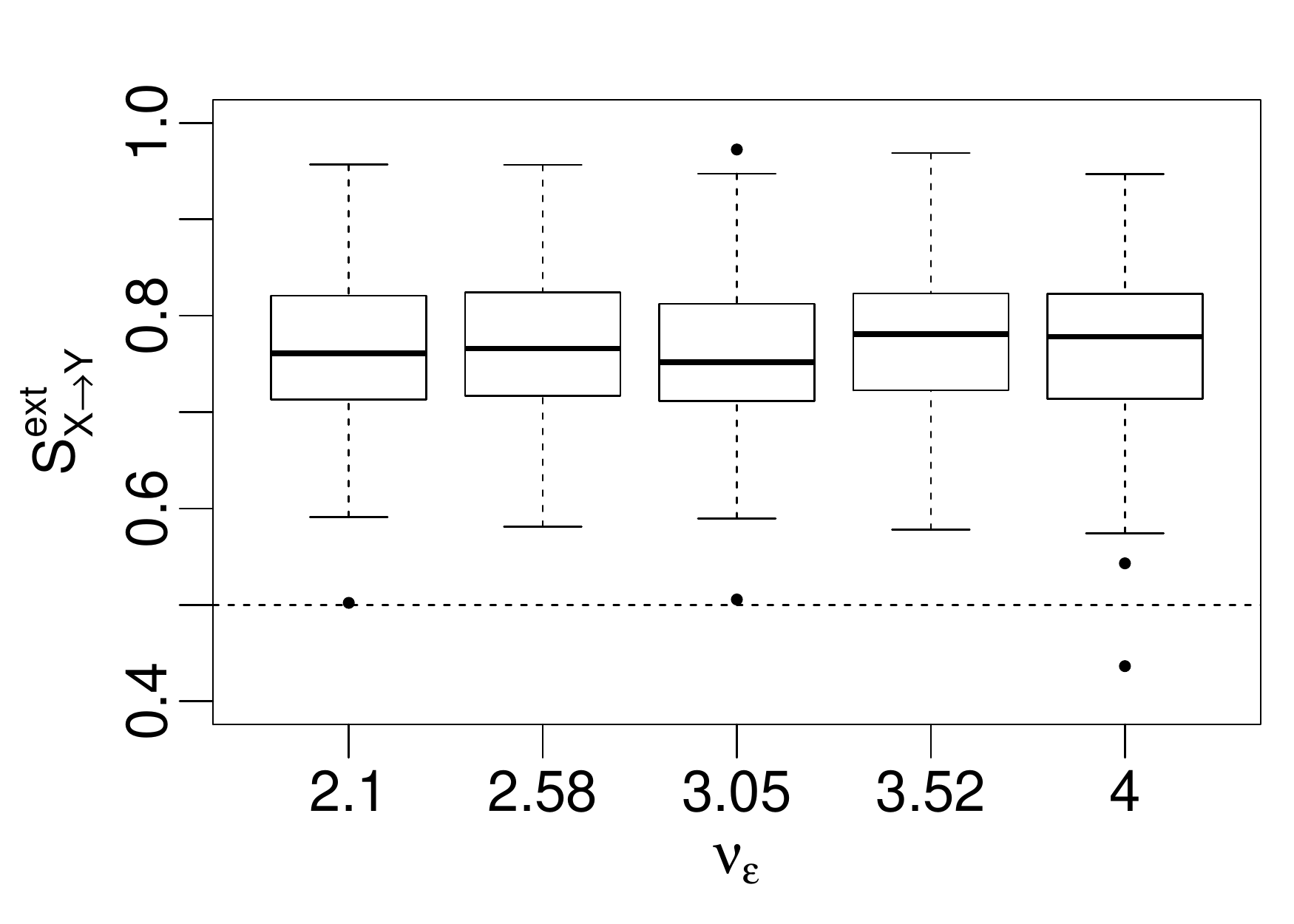}} \hfill
 \subfloat{\includegraphics[width=0.33\textwidth]{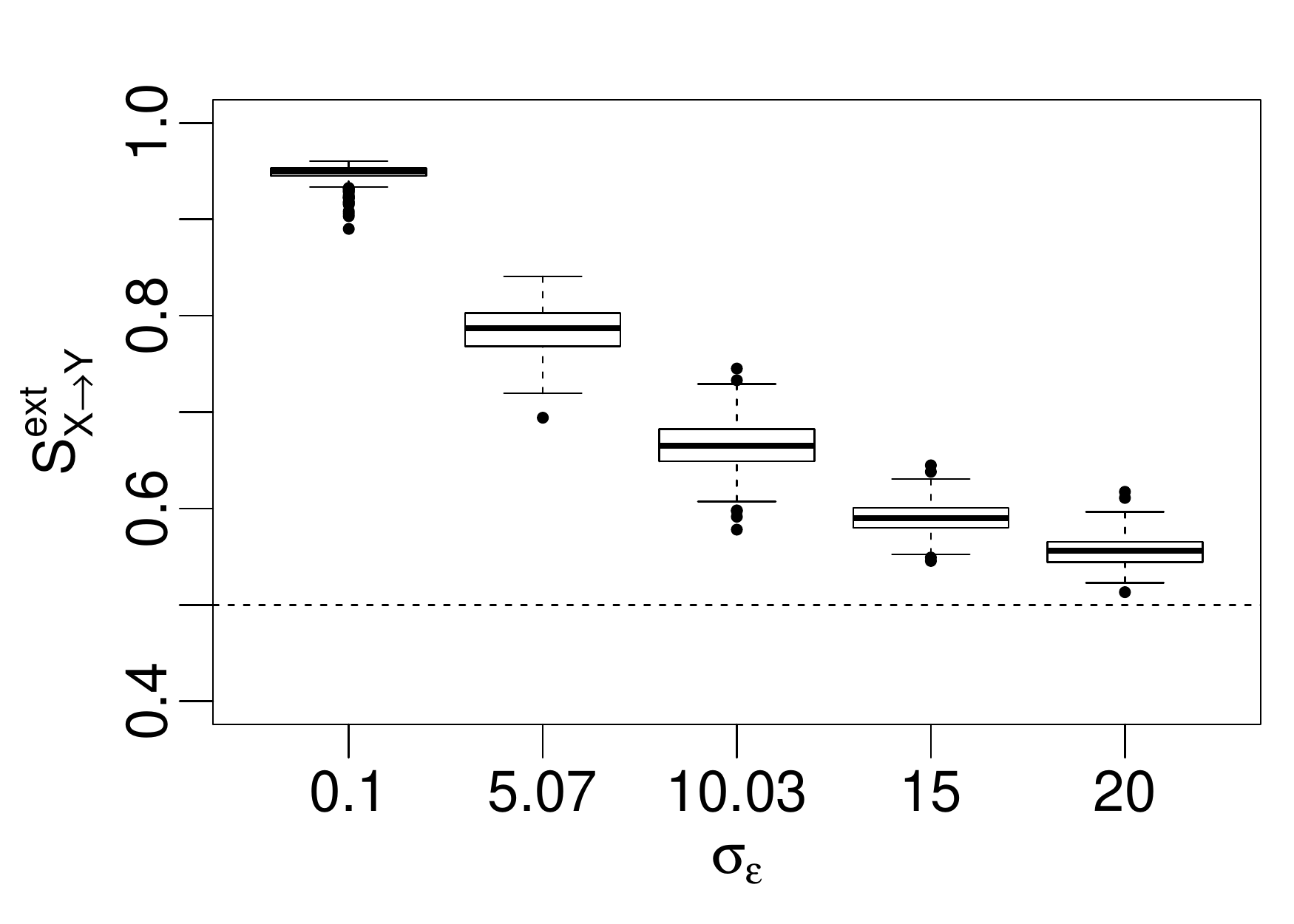}} \hfill
  \subfloat{\includegraphics[width=0.33\textwidth]{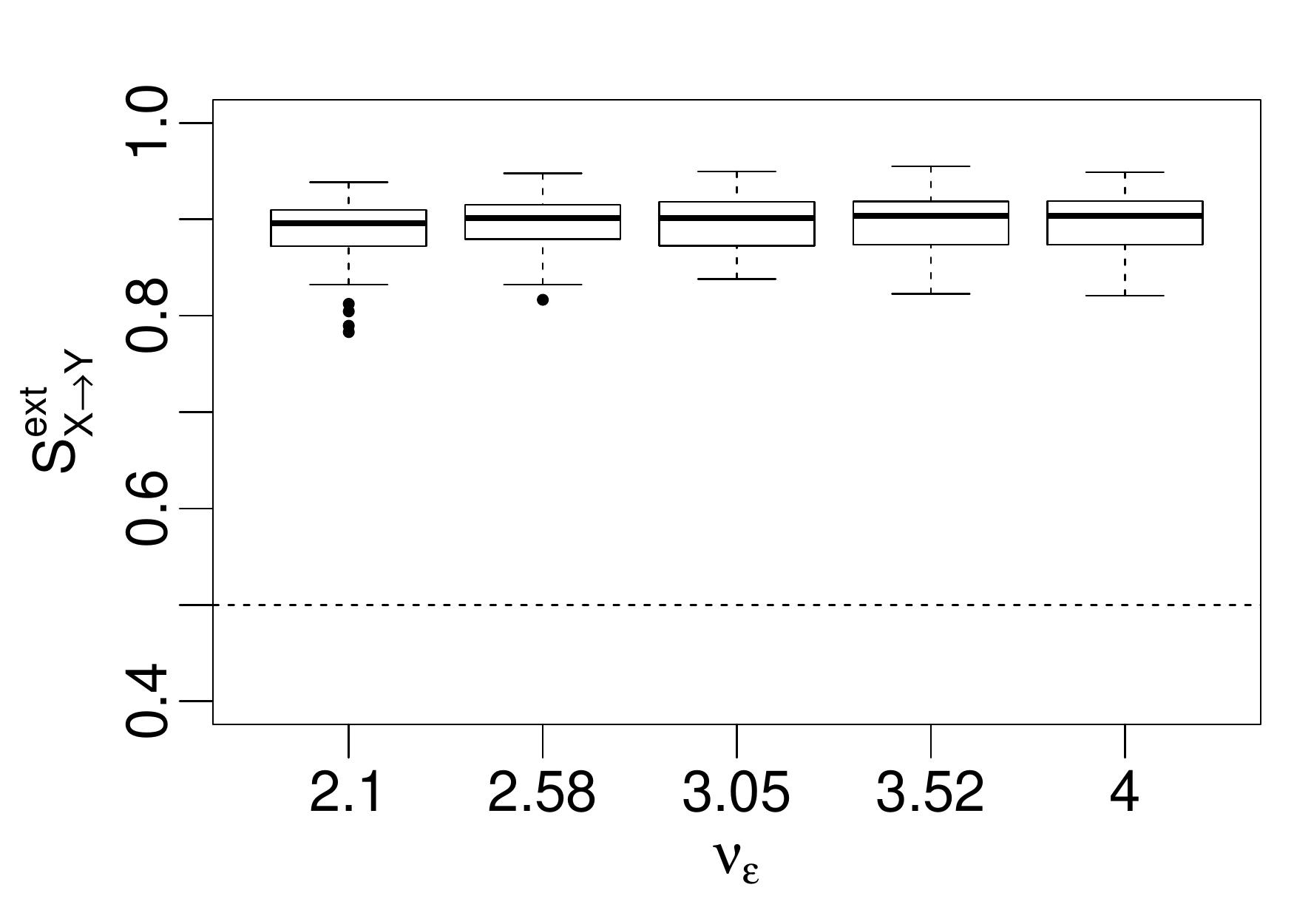}} \\
   \subfloat{\includegraphics[width=0.33\textwidth]{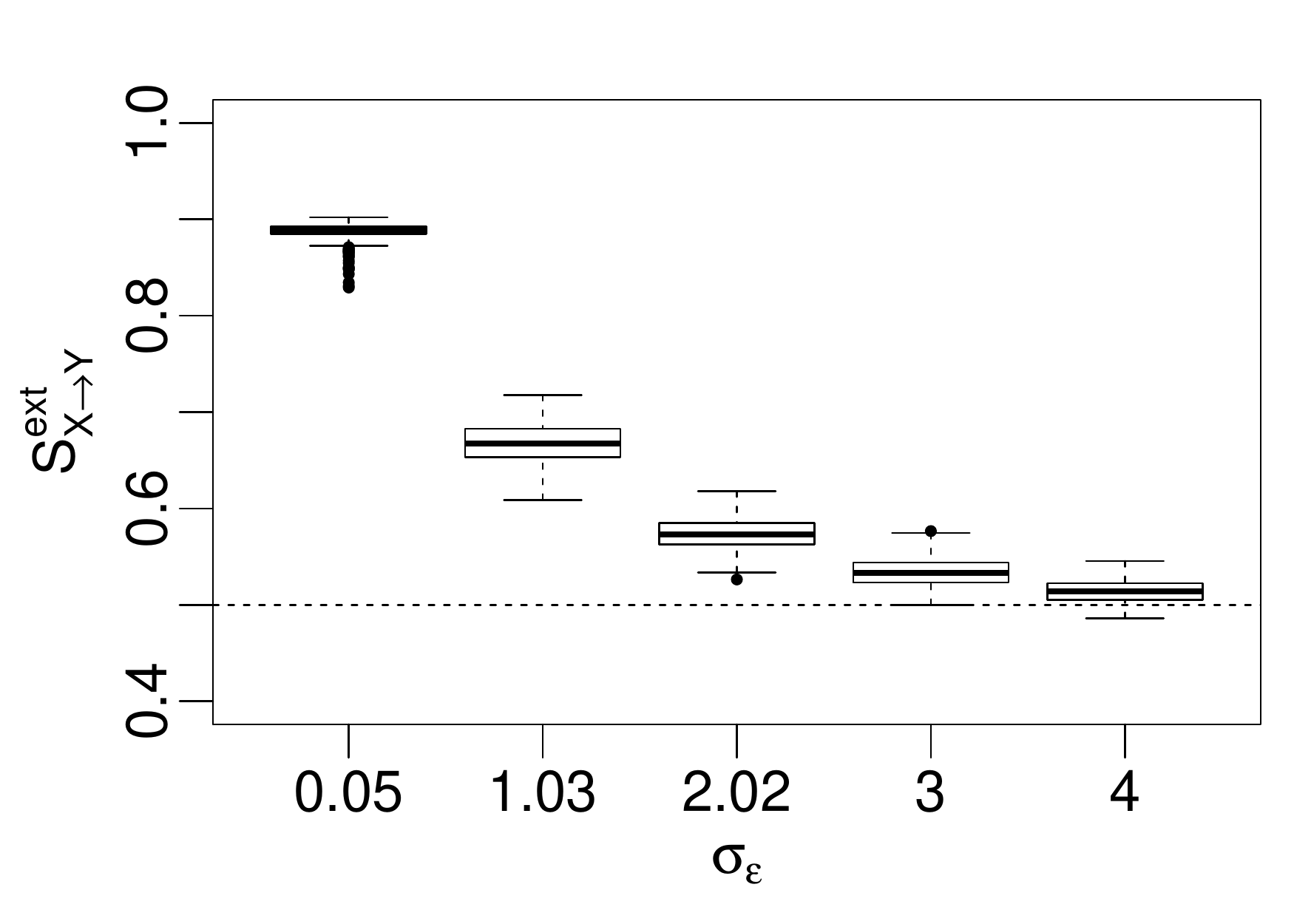}} \hfill
  \subfloat{\includegraphics[width=0.33\textwidth]{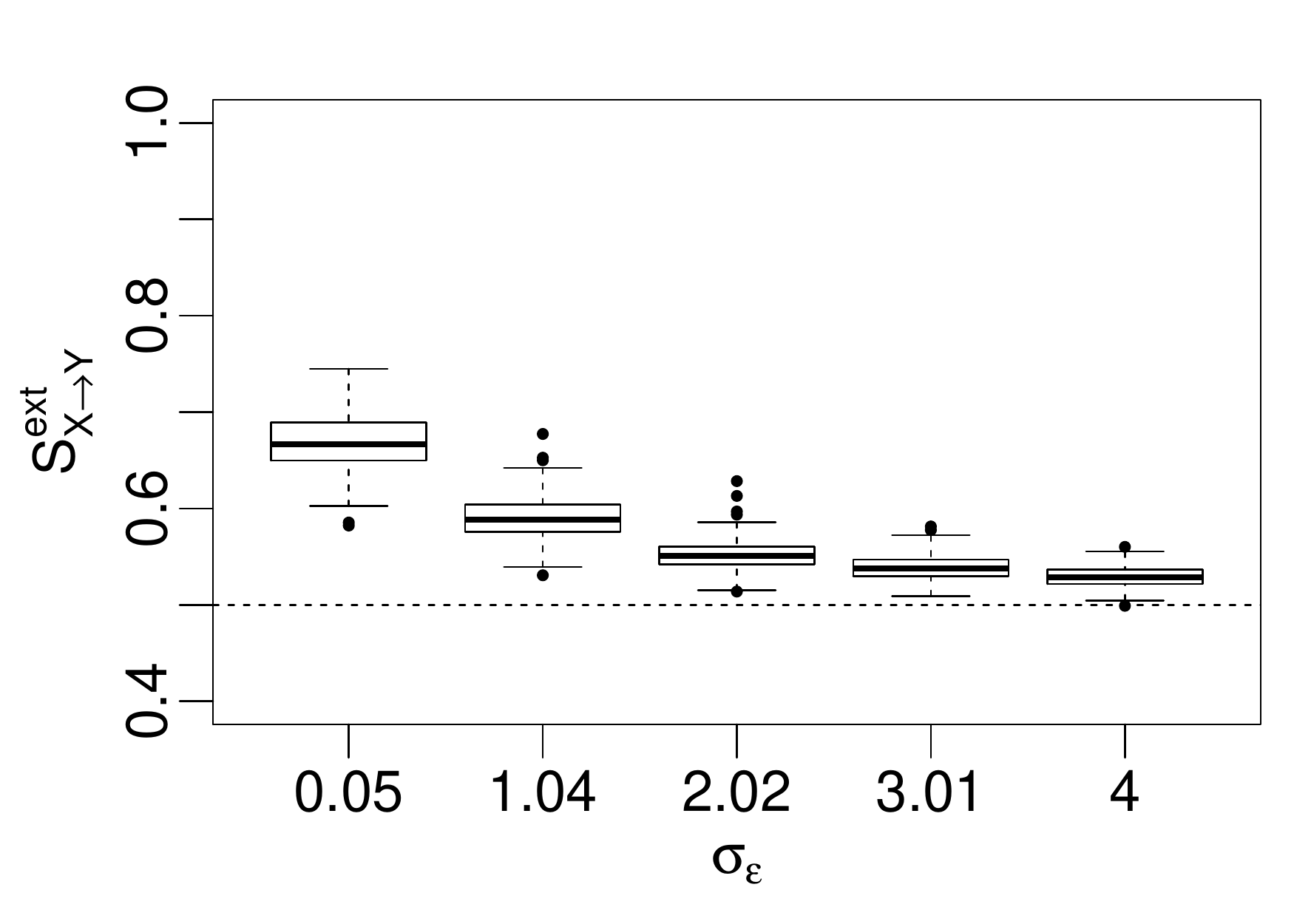}}
	\caption{Boxplots of the estimated score $S_{X \rightarrow Y}^{\text{ext}}$ for scenarios 1(a) (upper left panel); 1(b) (upper middle panel); 2(c) (upper right panel); 2(d) (lower left panel) and 3(e) (lower right panel). The $x$-axis represents either the degrees of freedom $\nu_{\epsilon}$ or the standard deviation $\sigma_{\epsilon}$.}
	\label{fig:S_anm}
\end{figure}

\bigskip
Figure~\ref{fig:S_anm} shows the estimated score $S_{X \rightarrow Y}^{\text{ext}}$
over the 300 simulations. In almost all cases, the scores are greater
than $0.5$ and CausEV can distinguish the 
cause from the effect in the extreme region of the data. The variability
of the noise affects the estimated score $S_{X \rightarrow Y}^{\text{ext}}$,
with larger variances inducing score values closer to $0.5$. For instance,
the scores for scenario 2(d) reach values smaller than $0.5$ when 
$\sigma_{\epsilon}=4$ and a decision on the causality in this noisy
scenario cannot be made. Heuristically, the value of the causal score reflects the strength of the causal signal, with decreasing causal scores approaching $0.5$ in the presence of increasing noise; see for instance the heuristic estimate of confidence defined by
\cite{Mooij2016} as the difference between causal scores of both directions.

%The decrease of the causal score in the presence
%of increasing noise is expected if one describes the evidence of
%causality using the heuristic estimate of confidence defined by
%\cite{Mooij2016} as $S_{Y \rightarrow X}^{\text{ext}}-S_{X \rightarrow Y}^{\text{ext}}$.

%\subsubsection*{\textsf{Comparison to methods for the mean level}}
Although numerous procedures have been proposed to infer the causal
direction from bivariate joint observational distributions, these methods
are not tailored to deal with exceedances in the joint upper tails
and do not exploit the asymptotically motivated results of the multivariate
extreme value theory. Moreover, when the causal mechanism is believed to be different in the tails than in the bulk of the distribution (see, e.g., \citet{Barbero2018}), the outcome of these methods might be affected by the causal relations in a moderate regime, thus misrepresenting these relations in an extreme regime. Therefore, to obtain a fair comparison, we apply four methods for the mean level
to the extreme set $\{(X^{\text{ext}}_i,Y^{\text{ext}}_i)\}_{i=1}^{n_u}$ directly. 

\bigskip
The first method, LINGAM (Linear Non-Gaussian Acyclic Model) \citep{Shimizu2006}, assumes that the data generating process is linear with non-Gaussian noise and that there are no unobserved confounders. Although the assumption of linearity is violated in our simulation settings, a few scenarios, namely 2(c), 2(d), and 3(e), exhibit a causal relationship close to linearity in the upper tails of the cause $X$.
%Although the assumption of linearity is violated in our simulation scenarios, we expect a causal relationship close to linearity for large values of $X$ and $Y$ as it is dictated, in the upper quadrant region, by the asymptotic behaviour of the structural function $h$. 
The second method, IGCI (Information-Geometric Causal Inference) \citep{Janzing2012}, assumes the absence of the noise in a structural causal model between the cause and the effect, and uses Postulate~\ref{postulate1} to construct a score based on the Kullback--Leibler divergence between the densities of the cause and the effect and a reference measure. We choose the Gaussian reference measure as it has been shown by \cite{Mooij2016} to yield higher performances of the method than the uniform reference measure. Despite the assumption of a noiseless deterministic relation, \cite{Janzing2012} discuss the robustness of their method in a noisy regime such as \eqref{AN_structure}. The third method, CAM (Causal Additive Model), assumes additive noise structure between the effect and the cause with a Gaussian noise variable. This method is robust against misspecification of the noise distribution \citep{Buhlmann2014}. The fourth method, RESIT (Regression with Subsequent Independent Test), assumes a structural equation model between $X$ and $Y$ with additive noise and performs a HSIC test \citep{Gretton_2008} for independence between the effect and the residuals of a generalized additive model of the effect as a function of the cause. The assumptions of the latter two methods are not violated in our simulation scenarios. 

\begin{figure}[!h]
\centering 
 \subfloat{\includegraphics[width=0.33\textwidth]{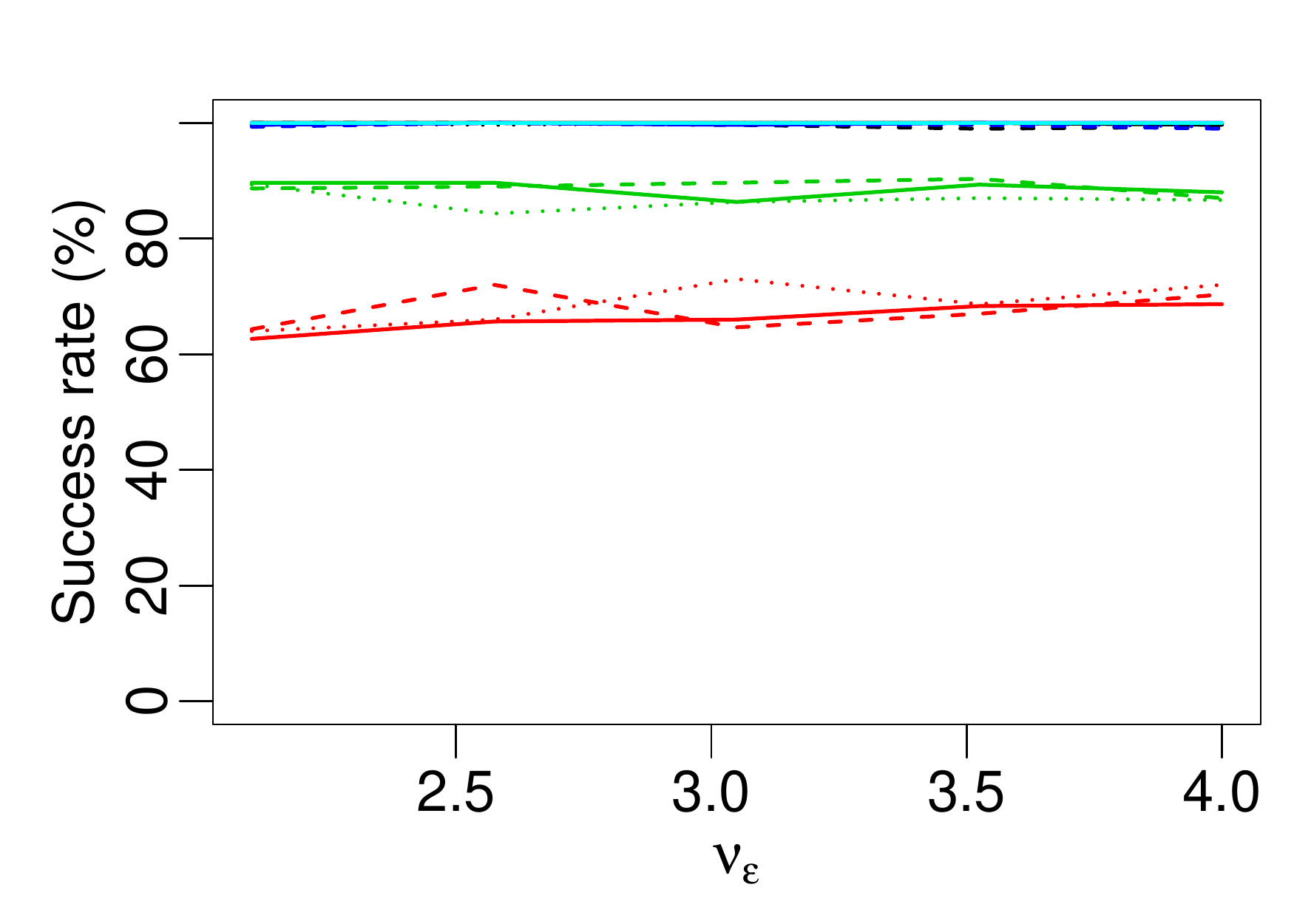}} \hfill
 \subfloat{\includegraphics[width=0.33\textwidth]{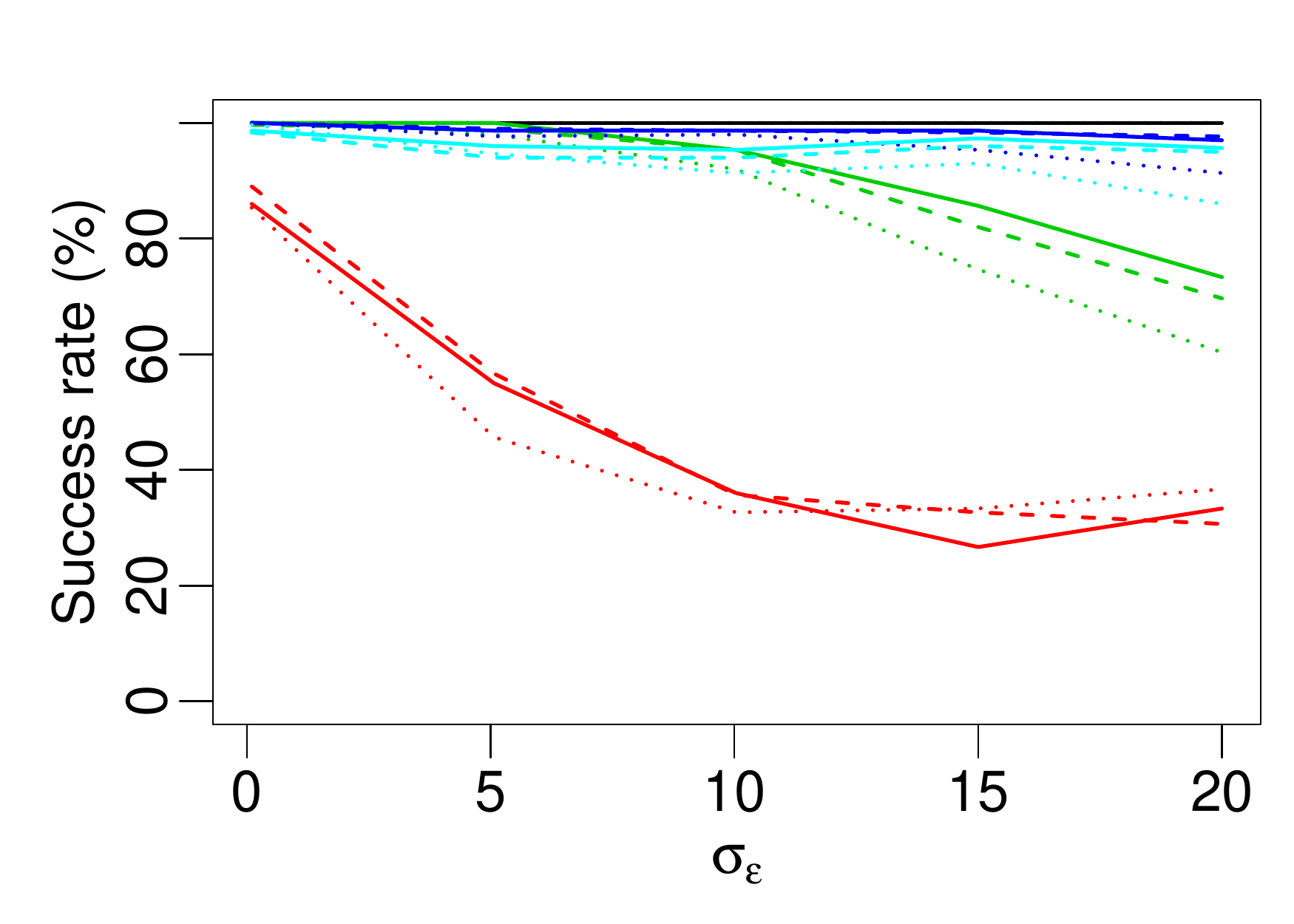}} \hfill
  \subfloat{\includegraphics[width=0.33\textwidth]{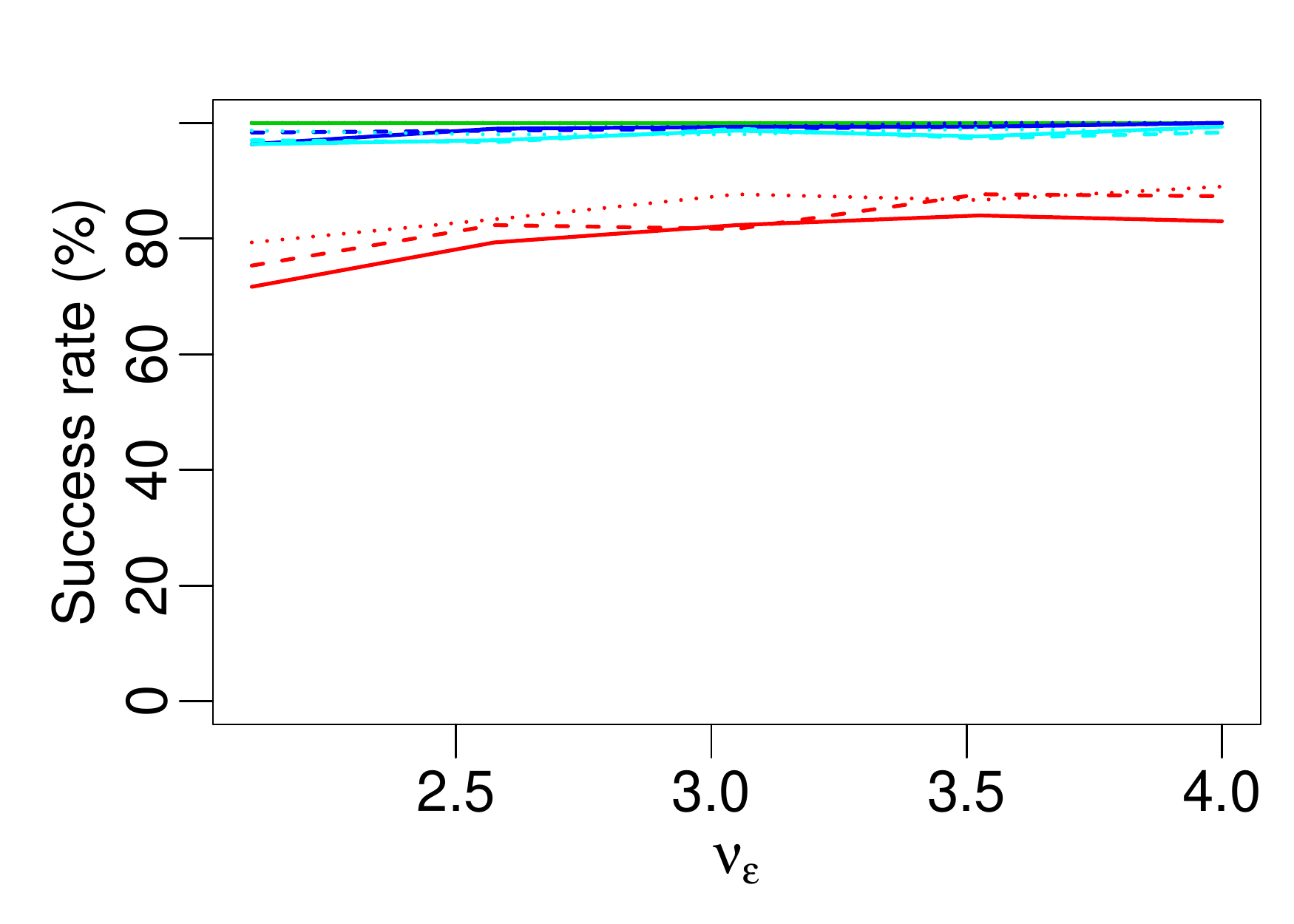}} \\
   \subfloat{\includegraphics[width=0.33\textwidth]{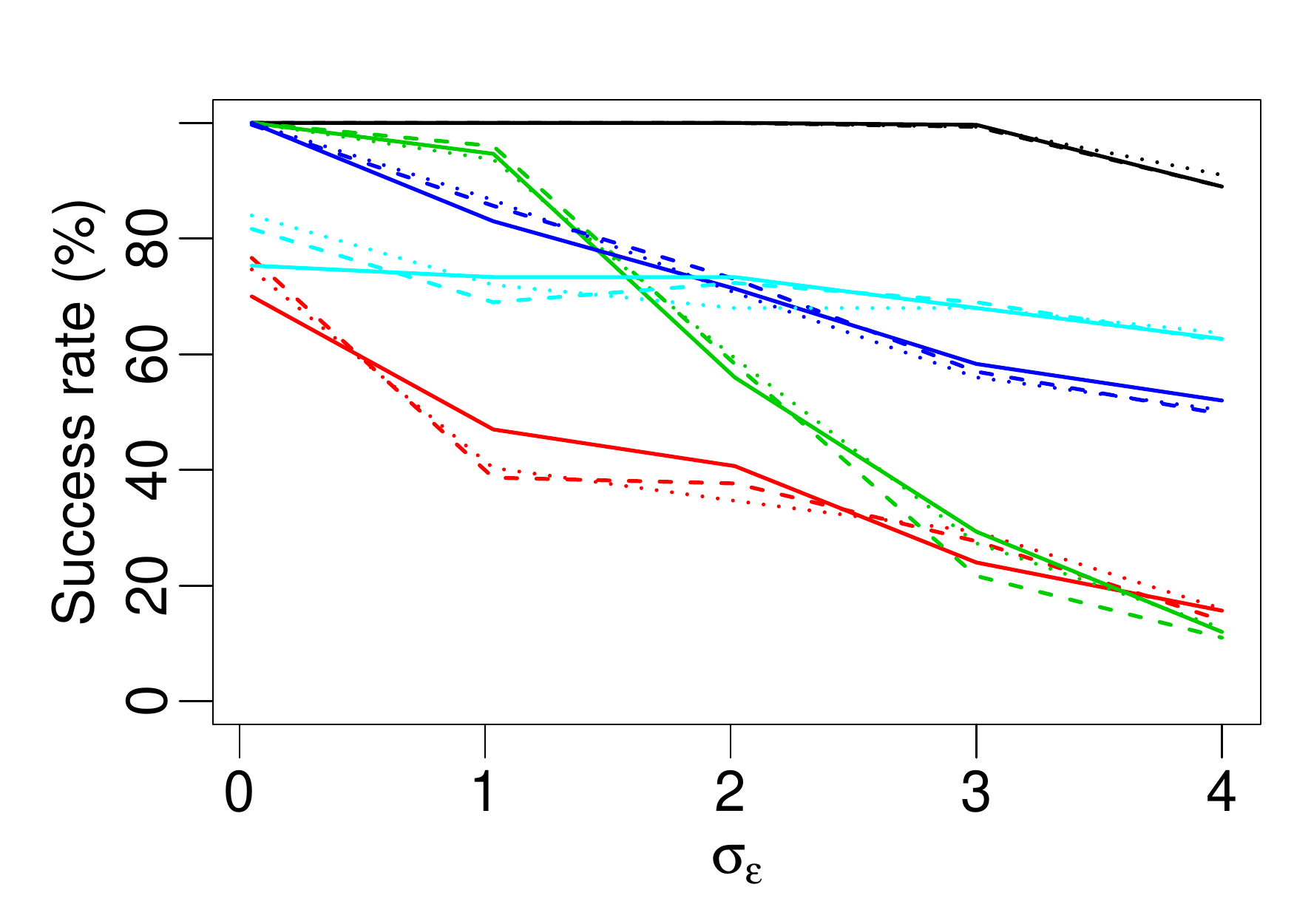}} \hfill
  \subfloat{\includegraphics[width=0.33\textwidth]{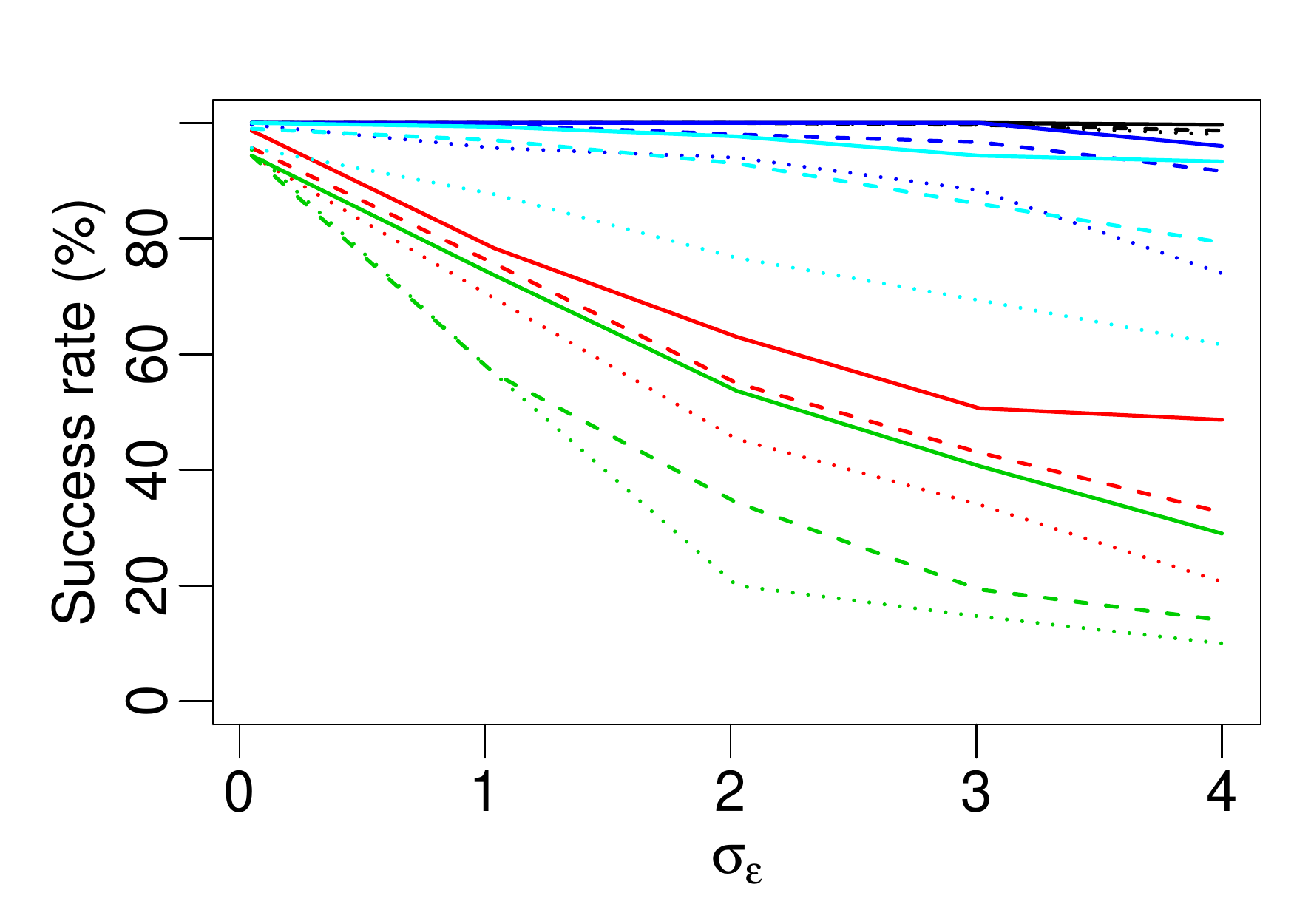}}
	\caption{Success rates of the bivariate causal discovery methods as a function of the degree of freedom $\nu_{\epsilon}$ or the standard deviation $\sigma_{\epsilon}$ of the noise. Panels correspond to the scenarios as described in Figure~\ref{fig:S_anm}. Results of CausEV (black), LINGAM (green), IGCI (light blue), CAM (dark blue), and RESIT (red) are represented in solid ($u=0.95$), dashed ($u=0.93$), and dotted ($u=0.9$) lines.}
	\label{fig:compet_method_anm}
\end{figure}

We compute the success rate, i.e., the percentage of repetitions (out of 300) inferring the true causal direction. Additionally, we assess the sensitivity of the success rate to the threshold choice by considering joint exceedances in the upper quadrant $\{(X^{\text{ext}}_i,Y^{\text{ext}}_i)\}_{i=1}^{n_u} = 
\{ (X_i,Y_i) : X_i>F^{-1}_X(u) \ \text{and } Y_i>F^{-1}_Y(u)\}$, with $u=0.9, 0.93, 0.95$ and fixed size $n_u=55$. Figure~\ref{fig:compet_method_anm} shows the results for all methods and at all considered marginal thresholds. CausEV is the clear winner as expected and its performance is unaffected by the increasing bias resulting from applying asymptotic extreme value models at lower thresholds.
The performances
of the methods CAM and IGCI are comparable in some
settings, but deteriorate with increasing noise variance. Lower marginal thresholds lead to decreasing success rates for the four mean-based methods in the Gaussian noise scenario 3(e). This is unsurprising since, by taking higher marginal thresholds, the causal structure is less spoiled by the light-tailed noise in these scenarios.

\subsection{\textsf{Robustness to the assumption of causality}}
Our use of the weaker version of Postulate~\ref{postulate2} given by
\eqref{algo_indep} assumes $X$ and $Y$ are causally related.
We now carry out experiments under three settings where this assumption
does not hold.

\begin{figure}[t]
	\centering 
 \subfloat{\includegraphics[width=0.33\textwidth]{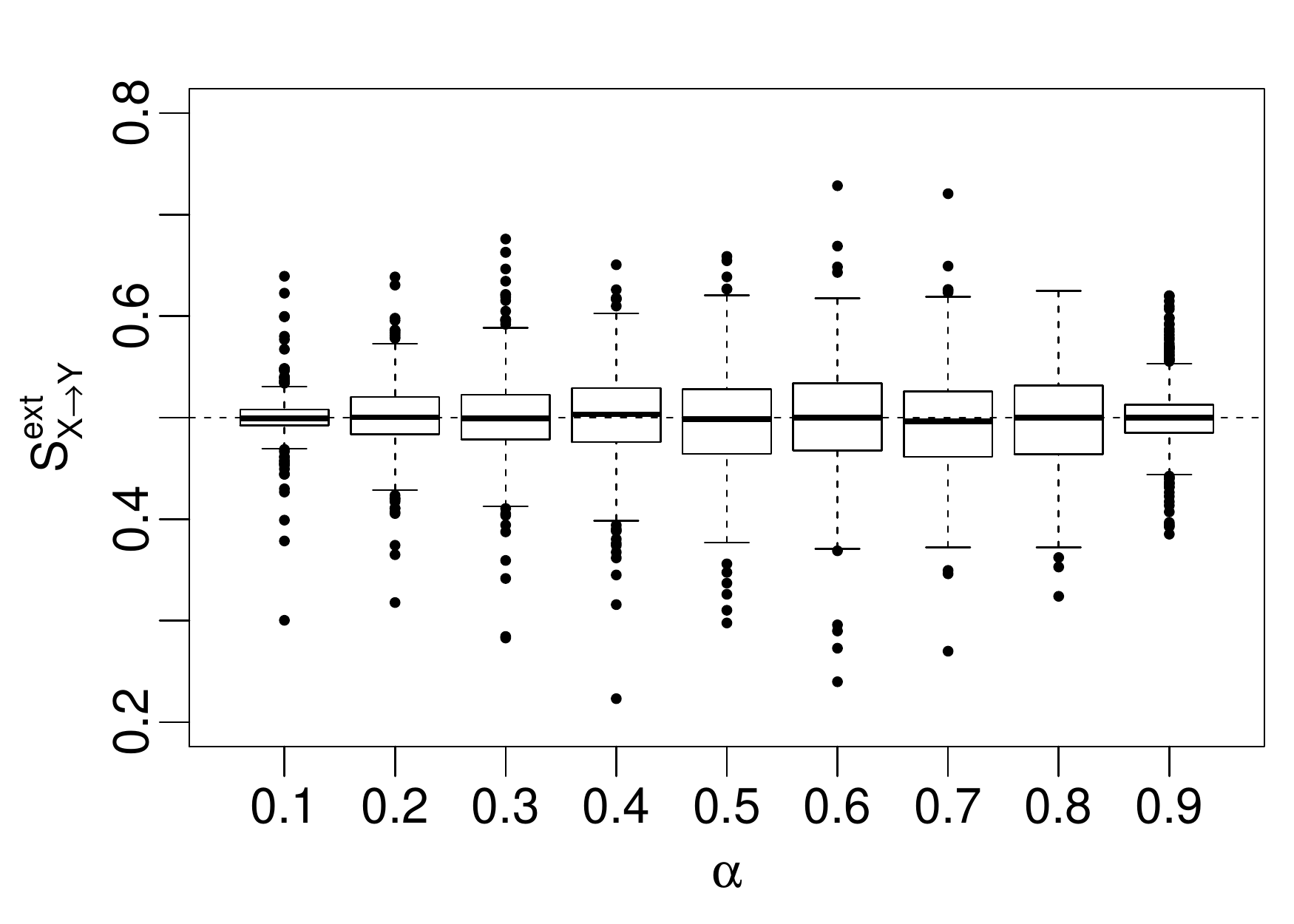}} \hfill
 \subfloat{\includegraphics[width=0.33\textwidth]{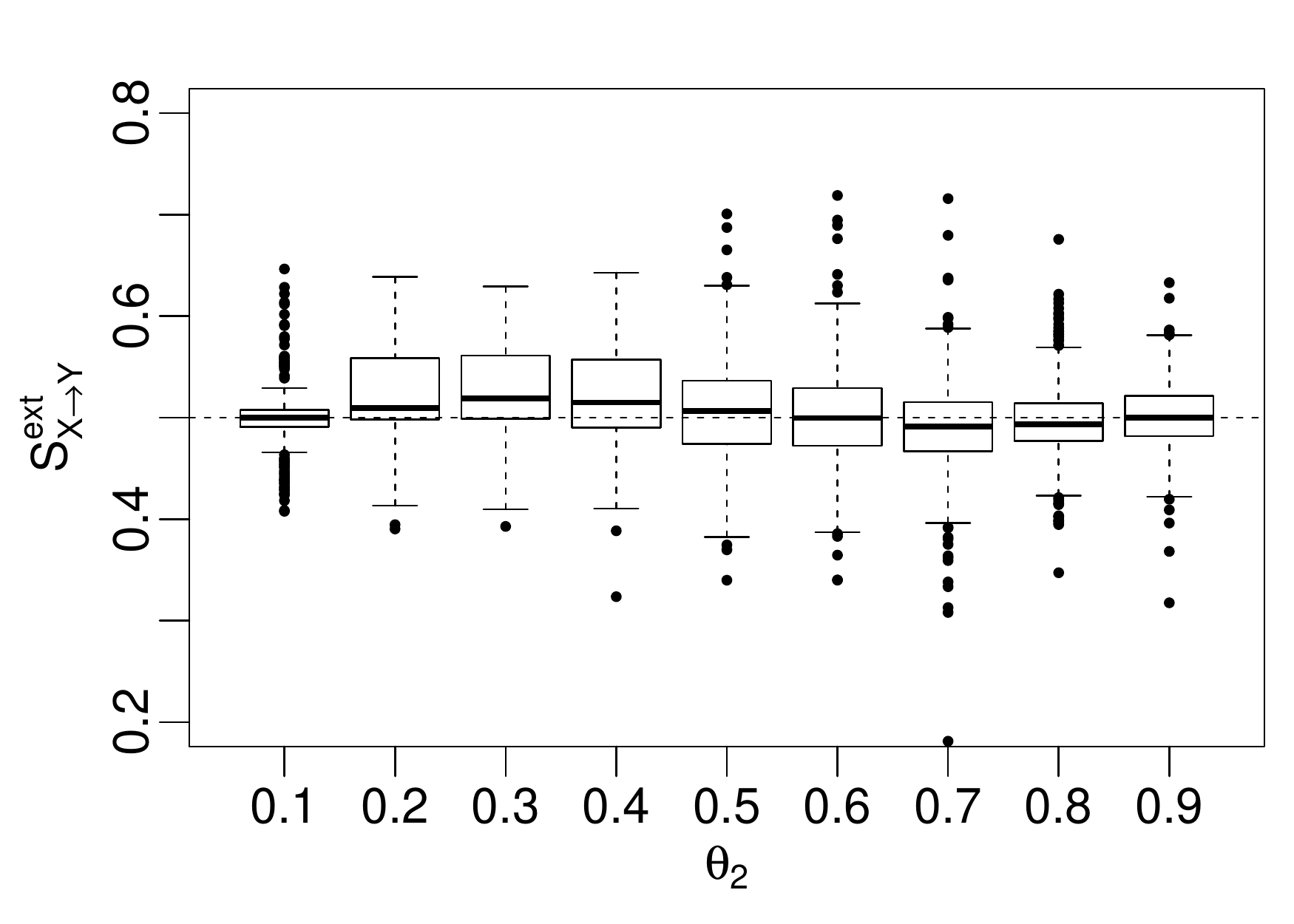}} \hfill
 \subfloat{\includegraphics[width=0.33\textwidth]{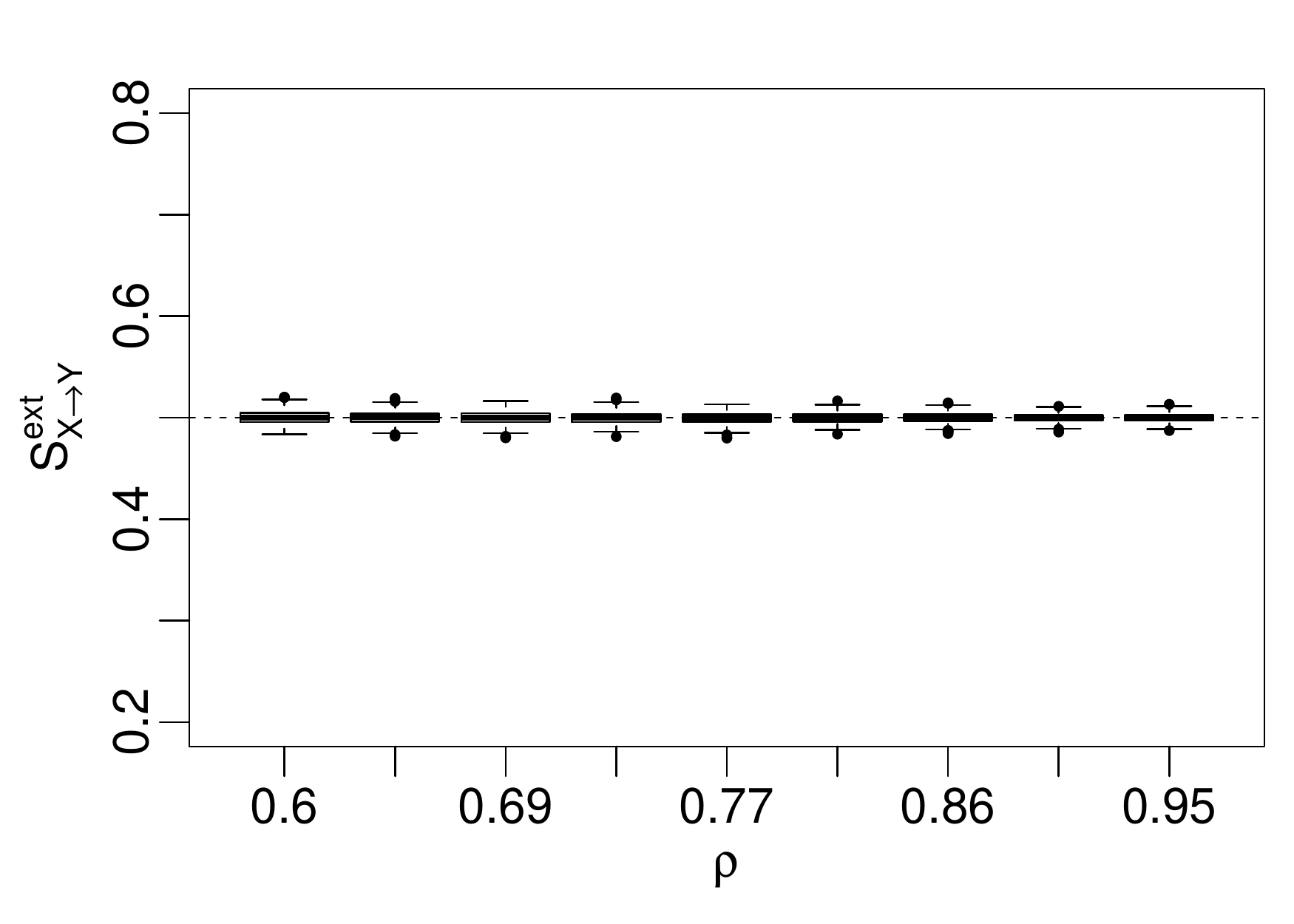}}
	\caption{Boxplots of the estimated score $S_{X \rightarrow Y}^{\text{ext}}$ in the case of bivariate generalized extreme value distributions with logistic copula (left panel) against the dependence parameter $\alpha$, and asymmetric logistic copula (middle) against the asymmetry parameter $\theta_2$, and in the case of a bivariate Gaussian distribution (right panel) against correlation $\rho$. }
	\label{fig:S_bivGev_Gauss}
\end{figure}

%\subsubsection*{\textsf{Data generated from an extreme value copula}}
Here, the MDL principle is applied to the correct class of models, i.e., the true underlying model belongs to the class of considered models. We simulate $(X,Y)$ from bivariate generalized extreme value distributions with the symmetric and asymmetric logistic copulas \cite{Tawn_1988}. Under the symmetric logistic setting, the strength of asymptotic dependence between $X$ and $Y$ depends on the parameter $\alpha \in (0,1]$, which we vary from $0.1$ (strong dependence) to $0.9$ (weak dependence). Two additional parameters $(\theta_1, \theta_2) \in [0,1]^2$ control the asymmetry of the dependence structure between $X$ and $Y$ in the asymmetric logistic case. Under this setting, we set the overall dependence parameter $\alpha=0.2$ and explore different levels of asymmetry by fixing $\theta_1=1$ and varying $\theta_2=0.1, \ldots,0.9$. The symmetric case is retrieved when $\theta_1=\theta_2=1$. By construction, both resulting random vectors satisfy the assumptions of our model class but neither $X$ causes $Y$ nor $Y$ causes $X$. We investigate the causality in the upper quadrant $\{(X^{\text{ext}}_i,Y^{\text{ext}}_i)\}_{i=1}^{n_u} = \{ (X_i,Y_i) : X_i>F^{-1}_X(0.95) \ \text{and } Y_i>F^{-1}_Y(0.95)\}$ and base the experiments on $300$ repetitions where we fix the number of concurrent exceedances at $n_u=55$. Figure~\ref{fig:S_bivGev_Gauss} shows the estimated scores $S_{X \rightarrow Y}^{ext}$ as a function of the logistic dependence parameter $\alpha$ (left panel) and the asymmetry parameter $\theta_2$ of the asymmetric logistic copula (middle panel). Based on the mean estimates of
the score $S_{X \rightarrow Y}^{ext}$, and for the different considered strengths of asymptotic dependence and asymmetry, CausEV is robust against the absence of a causal mechanism in the data and yields few false positives.

%\subsubsection*{\textsf{Data generated from an asymptotically independent copula}}
We now add model misspecification: we repeat the above simulation scheme but with a random vector $(X,Y)$ simulated from a bivariate Gaussian distribution with correlation $\rho$ varying from $0.6$ to $0.95$. The Gaussian copula is asymptotically independent, i.e., although we observe residual tail dependence at finite thresholds, the associated coefficient of tail dependence is $\chi=0$. Assuming an extreme value copula in the joint tail for $\{(X^{\text{ext}}_i,Y^{\text{ext}}_i)\}_{i=1}^{n_u} = \{ (X_i,Y_i) : X_i>F^{-1}_X(0.95) \ \text{and } Y_i>F^{-1}_Y(0.95)\}$ will overestimate the dependence in the joint tails and we assess here the resulting consequences on the causal score $S_{X \rightarrow Y}^{\text{ext}}$. Figure~\ref{fig:S_bivGev_Gauss} (right) shows
the estimated scores $S_{X \rightarrow Y}^{\text{ext}}$ as a function of $\rho$. The resulting mean estimates of
the score $S_{X \rightarrow Y}^{\text{ext}}$ show very few false positives, 
accurately detecting the absence of a causal mechanism in the data. 

\section{\textsf{Causal mechanism of extremes in the upper Danube basin}}
\label{danube}
The upper Danube basin covers most of the German state of Bavaria and
parts of Baden-W\"urtemberg, Austria and Switzerland. Frequent flooding
in the area has led to a well-developed system of gauging stations
covering the basin. Figure \ref{fig:topo_map} shows the locations of the
31 stations at which we have daily measurements of river discharge ($m^3/s$).
The daily series have lengths from 51 to 110 years and
are made available by the Bavarian Environmental Agency 
(\url{http://www.gkd.bayern.de}). There are 51 years of data for all 
stations from 1960--2010 and these data are analysed in 
\cite{AsadiDavisonEngleke2016} and \cite{Engelke_Hitz} where the
multivariate approaches require observations at all gauging stations
to be available.
We also consider data from 1960--2010 and
remove seasonality and trend issues following these authors: only
data for the months of June, July, and August are retained.
The most extreme precipitations and floods occur during summer 
\citep{Jeneiova2016}, see \cite{AsadiDavisonEngleke2016} for further
justification of temporal stationarity over the period retained.

\bigskip
Our interest lies in the causal structure between the peak flows at the 
31 stations.
Temporal dependence causes extreme discharges at a given station to 
occur in clusters. Extreme discharges at upstream stations may also cause
extreme discharges at downstream stations some days later. All these
extremes should be considered part of the same event and treated
as dependent. The data must thus be declustered and we do so
following \cite{AsadiDavisonEngleke2016}[Section 5.2].

\bigskip
Declustering yields approximately independent threshold exceedances 
at each station, and allows us to unveil the mechanism of causality in 
the extremes due to the inherent physical river dynamics, as opposed to the temporal causality in the discharges at flow-connected
stations that one would observe due to time lags allowing water propagation
through the network. According to \cite{Serinaldi2018}, time lags between peaks of flood events reflect flood duration which, for European summer flood events rarely exceeds the $9$-day period used by \cite{AsadiDavisonEngleke2016}. Starting from the $51 \times 92$ daily observations,
the declustering step results in $n=428$ independent events at all of the $31$ gauging stations. Each independent flood event represents a $31$-dimensional vector whose $i$-th entry corresponds to the maximum water discharge at the $i$-th station, observed within a $9$-day window where at least one station witnessed a large discharge value.

\begin{figure}[!h]
	\centering 
  \includegraphics[width=0.4\textwidth]{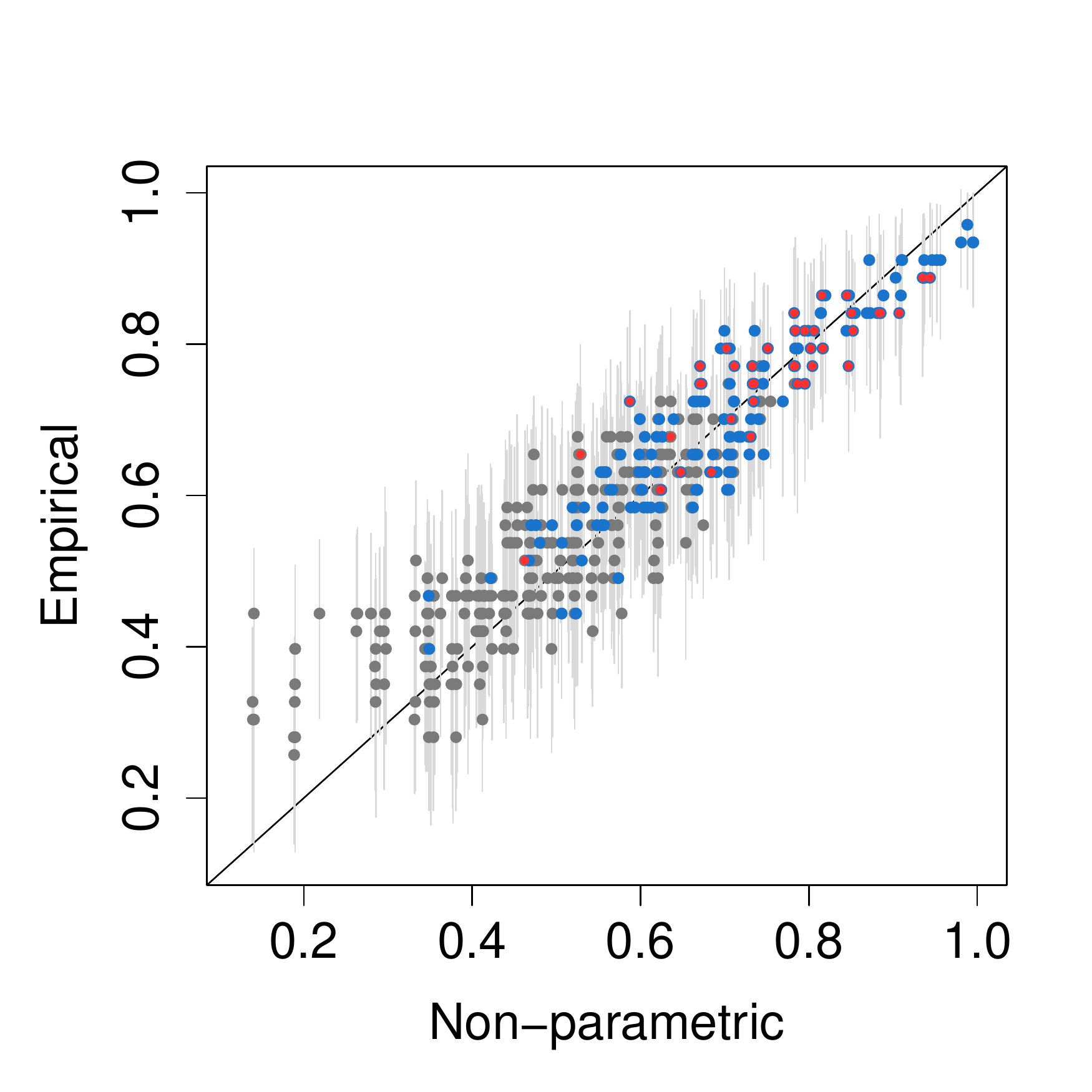}
	\caption{Empirical estimates of the $\chi$ coefficients along with $95\%$ bootstrap bounds, against estimated coefficients using the min-projection of \cite{Mhalla_OT_CD}. Estimates for flow-connected sites are represented in blue and estimates for causally-connected sites are in red.} 
	\label{fig:river_chi}
\end{figure}

\begin{figure}[!h]
	\centering 
 \subfloat{\includegraphics[width=0.49\textwidth]{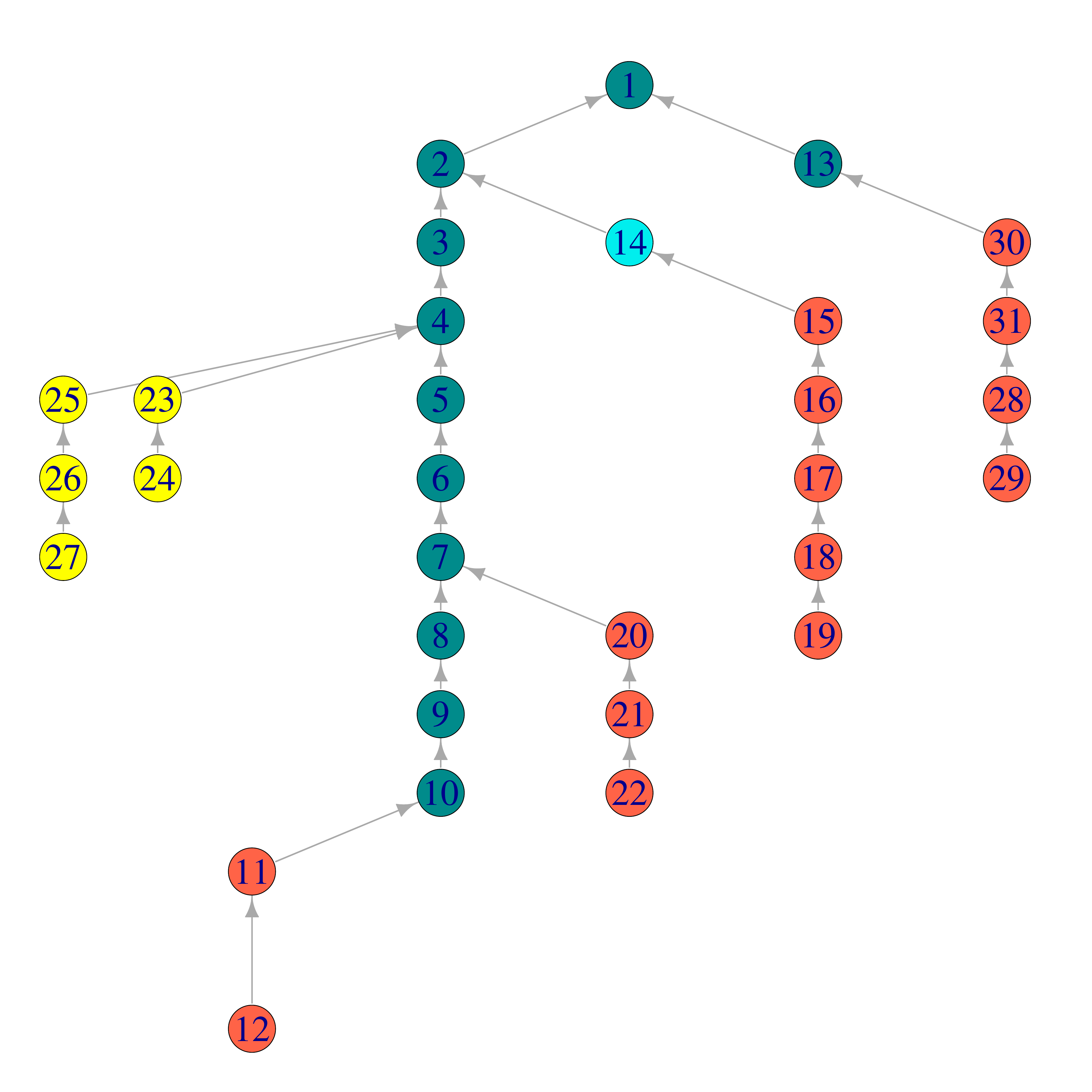}} \
 \subfloat{\includegraphics[width=0.49\textwidth]{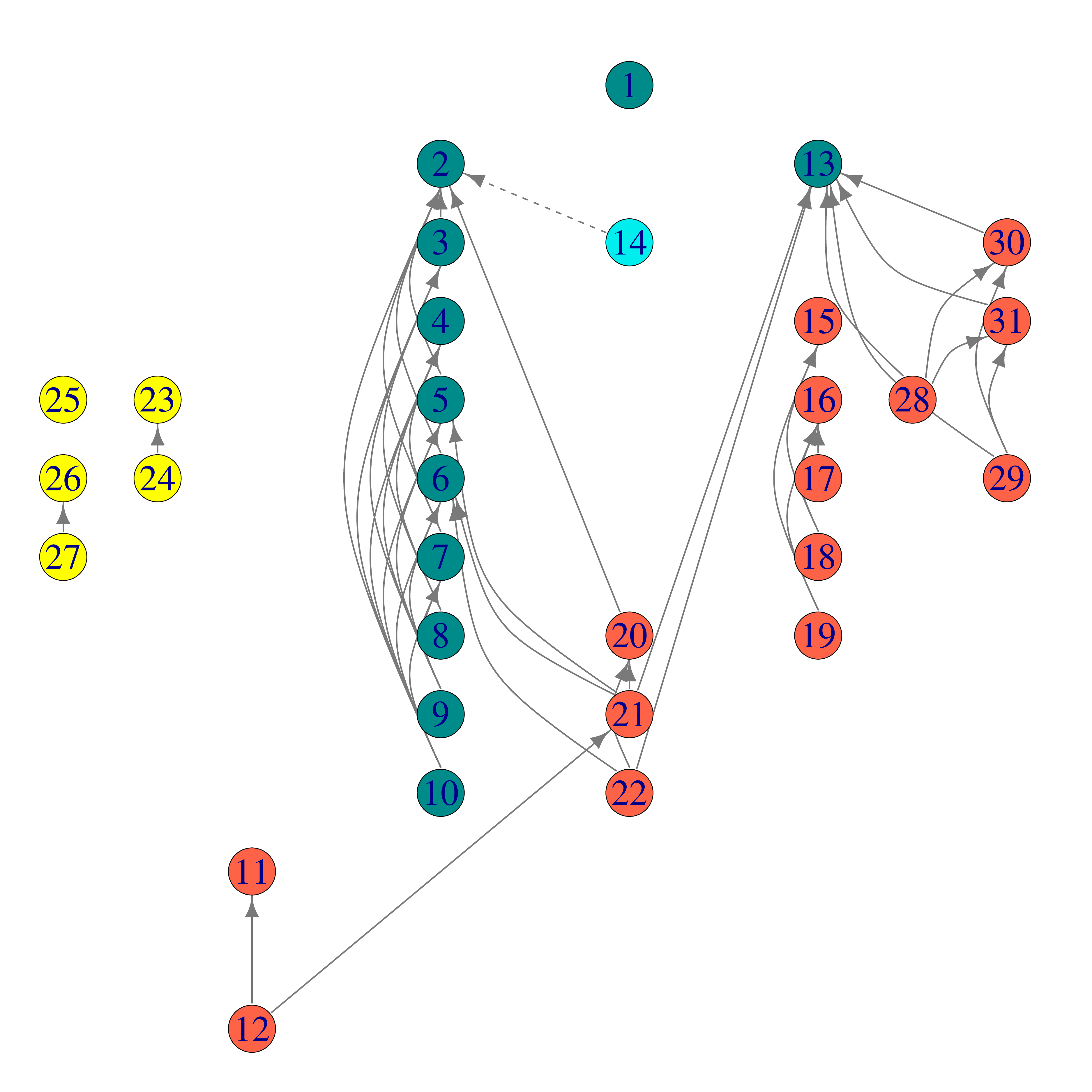}}
	\caption{Tree induced by water flow connections for the gauging stations in the Bavarian Danube basin (left) and oriented graph induced by the significant causal connections (right). Represented in dark green are the stations located on the Danube, in red the stations located on Alpine tributaries, in yellow the stations located on northern (to the Danube) tributaries, and in light blue the station located at the confluence between the Isar and the Danube.}
	\label{fig:tree_flows}
\end{figure}

\bigskip
We assess the validity of the assumption of asymptotic dependence in the data by computing empirical estimates of the coefficient of tail dependence $\chi$ \eqref{chi_formula} for all 465 possible pairs of stations. Figure~\ref{fig:river_chi} depicts the estimates across all pairs, along with $95\%$ block bootstrap uncertainty bounds. 
All estimates are strictly positive and we proceed by assuming asymptotic dependence in the data. This assumption is also justified by 
the exploratory analysis of \citet[Section 5.4]{AsadiDavisonEngleke2016} and the hydrological properties of the flow connections in the upper Danube basin; see for instance the tree induced by flow-connections in \cite{Engelke_Hitz},
reproduced in the left panel of Figure~\ref{fig:tree_flows}.

We consider the set of all possible pairs of stations in the river network
to which we apply CausEV. For each pair, we set the marginal thresholds at the $90\%$ empirical
quantiles leading to an average of $n_u=25$ joint threshold exceedances.
For the estimation of the extreme value copula, we transform the data 
empirically to the unit Fr\'echet scale and perform the min-projection of
\cite{Mhalla_OT_CD} at $500$ unequally spaced values in the unit simplex
$[0,1]$. The Pickands dependence function is then estimated based on the
deficits of the $10\%$ quantile of the min-projected random variable and
regularised using the COBS procedure; see \cite{Maechler} and \cite{Mhalla2017} 
for details. Figure~\ref{fig:river_chi} displays the resulting estimates 
of the coefficient of upper tail dependence for all pairs of stations $(X_i,X_j)$. Although this non-parametric estimation procedure might result in biased estimates as it assumes the validity of the extreme value copula \citep{Serinaldi2015}, it yields sensible estimates of
the tail correlation in the pairs, with slightly less
bias for the pairs of flow-connected sites.

\bigskip
The right panel of Figure~\ref{fig:tree_flows} displays, with solid edges,
the oriented graph induced by the pairs of stations with a significant causal
relation, i.e.\ with $95\%$ percentile bootstrap bounds 
for the causal score not containing the critical value $0.5$.
Bootstrap bounds are based on a yearly block bootstrap procedure applied
to the declustered data 300 times.
Forty-three oriented edges, from the cause to the effect,
are observed over the upper Danube basin. All follow the natural water flow topography. 
%\textcolor{green} {and the interpretation of these causal edges should be made in conjunction with the physical dynamics of the river network}.

\bigskip
Among the $43$ significant edges, we observe $28$ edges oriented towards stations located over the Danube river (dark green stations in Figure~\ref{fig:tree_flows}). Aside from the edges connecting stations along the Danube river and for which the direction agrees with the natural water flow, we uncover causal relations between the extreme discharges of stations located in the Alpine tributaries (red stations in Figure~\ref{fig:tree_flows}) of Lech (stations $20$--$22$) and Salzach (stations $28$--$31$) and the extreme discharges of stations located in the Bavarian Danube. 
These findings agree with the assertion of \cite{Skublics2016} and 
\cite{Bloschl2013} that summer floods in the Bavarian Danube basin are strongly caused by topographically enhanced precipitation at the northern rim of the Alps. 

\bigskip
We do not find edges in the northern tributaries of Regen 
and Naab (yellow stations in Figure~\ref{fig:tree_flows}) and
gauging station $4$, even if they are flow-connected.
This is in fact as expected since we are
only examining extreme summer events where causal signals between these tributaries and the Danube might be weak. According to \cite{Skublics2016}, it is warm winter snow melt and rainfall in these northern tributaries 
that lead to extreme discharges and winter flooding in the Bavarian
Danube basin.

\bigskip
We do not find an edge between the flow-connected gauging stations
$15$ and $14$. Looking at the daily water discharges at these stations,
we notice a strong linear correlation in the observations that
carries over in the tails with an empirical estimate of $\chi_{0.9}$ 
evaluated approximately at $0.89$. The observed linearity in the data
renders the causal discovery task infeasible as both causal directions
could be observed, although in reality only the one given by the flow
direction is possible.
%\begin{figure}[!h]
%	\centering 
% \subfloat{\includegraphics[width=0.45\textwidth]{Figures/Exploratory_Analysis/sites_15_14_1991_1.pdf}} \
%  \subfloat{\includegraphics[width=0.45\textwidth]{Figures/Exploratory_Analysis/sites_15_14_1991_2.pdf}}
%	\caption{Water discharges at the gauging sites $15$ (left) and $14$ (right) during summer 1991. Observations exceeding the $90\%$ empirical quantile, in each time series, are depicted in red.}
%	\label{site_14_15}
%\end{figure}

\bigskip
We find an edge between the Alpine tributaries Iller (station $12$)
and Lech (station $21$). These are not flow-connected,
but they originate in the same region of the Bregenz Forest Mountains
in the Alps, where common topographic effects induce moisture convergence during summer that triggers intense convective downpours \citep{Beniston2007}. The edge should be interpreted
cautiously, as extreme discharges along these tributaries are likely 
to be triggered by a set of confounders related to their geographical
locations. For example, the largest river discharges at these 
two stations were observed in August 2005 when
western Tyrol and the south of Bavaria witnessed extensive 
precipitation and high antecedent soil moisture \citep{BLU2006}. The oriented edge between stations $12$ and $21$ remains even after removing the August 2005 flood event. The Salzach tributary (stations $28$ to $31$), although originating in
the Alps, is not causally related to these tributaries. This is 
expected from the topographic map of the upper Danube basin where
the Salzach catchment is separated from the sub-catchment of the
 other Alpine tributaries by the Inn river (not represented in our dataset).
For instance, the most extreme discharge observed at station $28$
occurred in August 1977 and it coincides with the 26th largest
discharge at station $21$.

\bigskip
An advantage of CausEV is that the analyses are done
pairwise and do not require complete data at all stations for 
data to be retained, as in a multivariate approach. 
We restricted our analyses to data over 1960--2010 only because
the upper Danube basin presents non-stationarity over
longer periods due to the construction of dams and hydropower plants 
along the Danube and its tributaries prior to 1960.
Only a few segments in the upper
Danube basin are uninterrupted by a dam, e.g., that stretching
from Straubing (upstream of the gauging station $3$) to Vilhofen
(downstream of the gauging station $2$), and the segment linking 
gauging station $14$ on the Isar to the Danube; 
see Map 23.1 in \cite{mauser2015regional}. When all the available 
observations at these stations are considered, i.e.\ from the summer 
of $1901$ for station $2$ and from the summer of $1926$ for 
stations $3$ and $14$, we uncover a causal effect for the 
flow-connected pairs $3\longrightarrow 2$ and $14\longrightarrow 2$ and, as one would expect, there are no cycles in the resulting graph.
%The mean bootstrap estimate of the causal score can be used as a heuristic assessment of the evidence of causality in the different pairs. For instance, and as depicted by the simulations studies, higher values of the score reflect a higher evidence of causality. Based on this heuristic measure, one can order the edges and keep only a certain amount of causal relationships.

%{\color{red}and, as one would expect, there are no cycles in the
%resulting graph.} {\color{red} maybe here we should say that because all the oriented edges respect the natural flow topography, there is no cycles in the estimated structure which is something expected for a graph to be a directed acyclic graph (DAG) so we would not need to cut some edges if we wanted to get a DAG, no? Linda: yes but we would need to explain what a DAG is and more importantly explain its role in causal discovery methods with multivariate or high dimensional data; which we can do but probably somewhere else no?.} 
\section{Conclusion}
\label{conclu}
\bigskip
The approach described in this paper is a first step towards the development of causal inference for extreme values, a new and promising line of research. The resulting method is based on the asymptotically justified arguments of extreme value theory lifting the restriction on the knowledge of the true distribution underlying the observations. In the context of its application to the Danube, CausEV requires no additional information such as the topology of the network or the distances between the pairs of stations. Other applications where underlying causal relationship between extremes matters are thus easily treated.
An important point not treated in this paper is the possible presence of confounders. Future work will aim to adapt the current method by including the effects of common causes at extreme levels. Specifically, one can compute proxies for the Kolmogorov complexity conditional on a set of potential confounders by modelling the influence of these confounders on the conditional quantiles, that is on the marginal distributions and the extreme value copula.

\section*{Acknowledgements}
The first author acknowledges the support of the Centre de Recherches Mathématiques and the Canadian Statistical Sciences Institute. Support of the Swiss National Science Foundation is gratefully acknowledged by the first and second authors. The second author acknowledges the financial support of the Forschungsinstitut fuer Mathematik (FIM) of the ETH Zurich. The third author acknowledges the support of the Natural Sciences and Engineering Research Council of Canada RGPIN-2016-04114 and the Fondation HEC. The authors would like to thank Anthony C. Davison for his comments and suggestions. The authors also thank the Associate Editor and two anonymous referees for comments that improved the presentation of the results.
\newpage
\bibliographystyle{rss}
\bibliography{mybib} 

\end{document}

%% file: short_bib.tex
 \let\oldthebibliography=\thebibliography
 \let\oldendthebibliography=\endthebibliography
 \renewenvironment{thebibliography}[1]{%
   \footnotesize
   \oldthebibliography{#1}%
   \setlength{\parskip}{0mm}%
   \setlength{\itemsep}{0mm}%
 }{\oldendthebibliography}